\documentclass[reqno]{amsart}
\usepackage{amsmath,amssymb,latexsym,amscd}
\usepackage{diagxy}
\newcommand{\beq }{\begin{equation}}
\newcommand{\eeq }{\end{equation}}
\newcommand{\e}{\epsilon}

\newcommand{\va}{\vartheta}

\newtheorem{theorem}{Theorem}
\newtheorem{lemma}{Lemma}
\newtheorem{proposition}{Proposition}
\newtheorem{corollary}{Corollary}
\newtheorem{definition}{Definition}
\theoremstyle{definition}
\newtheorem{remark}{Remark}
\newtheorem{example}{Example}
\numberwithin{equation}{section}

\begin{document}

\title[Variational Theory of Balance Systems]%
{\textbf{Variational Theory of Balance Systems}}
\author{Serge Preston}\address{Department of Mathematics and Statistics, Portland State University,
Portland, OR, U.S.}\email{serge@mth.pdx.edu}

\begin{abstract}
 Abstract. In this work we apply the Poincare-Cartan formalism of the Classical Field Theory to study the systems of balance equations (balance systems). We introduce the partial k-jet bundles $J^{k}_{p}(\pi)$ of the configurational bundle $\pi : Y \rightarrow X$ and study their basic properties: partial Cartan structure, prolongation of vector fields, etc. A constitutive relation $C$ of a balance system $BC$ is realized as a mapping between a (partial) k-jet bundle $J^{k}_{p}(\pi)$ and the extended dual bundle $\Lambda^{n+(n+1)}_{2/1}Y$ similar to the Legendre mapping of the Lagrangian Field Theory.\par
Invariant (variational) form of the balance system $\mathcal{B_C}$ corresponding to a constitutive relation $\mathcal{C}$ is studied. Special cases of balance systems -Lagrangian systems of order 1 with arbitrary sources and RET (Rational Extended Thermodynamics) systems are characterized in geometrical terms.
 Action of automorphisms of the bundle $\pi $ on the constitutive mappings $\mathcal{C}$ is studied and it is shown that the symmetry group $Sym(\mathcal{C})$ of $\mathcal{C}$ acts on the sheaf of solutions $Sol_\mathcal{C}$ of balance system $\mathcal{B_{C}}$. Suitable version of Noether Theorem for an action of a symmetry group is presented together with the special forms for semi-Lagrangian and RET balance systems and examples of energy momentum and gauge symmetries balance laws.

\end{abstract}
\maketitle
\keywords{ k-jet bundle, balance law, Poincare-Cartan form, Noether Theorem}
\vskip0.3cm
\today
\section{\textbf{Introduction.}}
Systems of balance equations (balance system, shortly, {\emph{BS}}) for the fields $y^\mu$ accompanied by the proper constitutive relations $\mathcal{C}$ are the basic tools of Continuum Thermodynamics.  In this work we develop a variational (in the sense of Poincare-Cartan form) theory of balance systems.\par
In Sec.2 we introduce basic notions: k-jet bundles, contact decomposition of forms, Poincare-Cartan formalism of order 1, balance systems.
In Sec.3 the partial jet bundles $J^{k}_{p}(\pi)$ of configurational bundles $\pi:Y^{n+m}\rightarrow X^n$ are introduced as the appropriate domains of the constitutive relations of the corresponding field theory. These bundles corresponds to an almost product structure on the base manifold $M$ (space-time decomposition is the main example). Partial Cartan structures and prolongation of vector fields from $Y$ to the partial jet bundles are studied.  In Sec.4 the constitutive relations $\mathcal{C}$ are defined as the mappings $C:J^{k}_{p}(\pi)\rightarrow \Lambda^{(n+(n+1)}_{2/1}Y$ to the  bundle $\Lambda^{(n+(n+1)}_{2/1}Y$ of  $n+(n+1)$-forms on $Y$ annulated by two $\pi$-vertical tangent vectors factorized by the bundles of similar forms annulated by one vertical tangent vector.  Lifting ${\hat C}: J^{k}_{p}(\pi)\rightarrow \Lambda^{(n+(n+1)}_{2}Y$ of such a mapping induces the $n+(n+1)$-form $\Theta_{\hat{C}}$ on the partial jet bundle - Poincare-Cartan form of considered field theory. Special forms and examples of such CR are presented: lifting covering constitutive relations, semi-Lagrangian, RET, relations defined by the Lagrangian and a Dissipative potential, etc. In Sec.5 the invariant form of a balance system corresponding to a constitutive relation $\mathcal{C}$ is introduced and studied.  Cases of Lagrangian field theories of the first order and of the Rational Extended Thermodynamics as a specific case of the balance systems are described. We study the admissible variations of the fields $y\mu$ splitting the invariant equation into separate balance equations. An appropriate modification of the source term allows us to remove conditions on the admissible variations. \par Action of automorphisms of configurational bundle $\pi:Y\rightarrow X$ on the constitutive relation $C$ is defined and its effect on the corresponding Poincare-Cartan form and on the sheaf of solutions of the balance system is studies is Sec. 6.\par  In Sec.7 we prove the appropriate version of the first Noether Theorem associating the new balance laws with an infinitesimal symmetries of a constitutive relation $\mathcal{C}$ and determine when such a balance law is actually, the conservation law.  Examples of the energy-momentum balance law and that of the gauge symmetries are presented. In Sec.8 we sketch the application of present scheme to the search and classification of the "secondary balance laws of a given \emph{BS}, including the entropy balance law. As an example, we present classification of secondary balance laws for the Cattaneo heat propagation balance system and the constitutive restrictions on the CR $\mathcal{C}$  that follows from the II law of thermodynamics. \par
Short exposition of main results of this paper was presented in the Conference Proceedings \cite{Pr3}.
\section{Settings and the framework}
\subsection{Notations and preliminaries}
\par
Throughout this paper $\pi:Y\rightarrow X$ will be a (\emph{configurational}) fibred bundle with a n-dim connected paracompact smooth ($C^\infty $)  manifold $X$ as the base and a total space $Y,\ dim(Y)=n+m$. Fiber of the bundle $\pi$ is a $m$-dim connected smooth manifold $U$.\par
Base manifold $X$ is endowed with a (pseudo)-Riemannian metric $G$.  Volume form of metric $G$ will be denoted by $\eta$. In this paper we will not be dealing with the boundary of a base manifold  $X$,  in applications $X$ can be considered as an open subset of $R^n$ or as a compact manifold. As an basic example, we consider the case where $X=T\times B$ is the product of time axis $T$ and an open material manifold (or a domain in the physical space) $B$. \par

We will be using fibred charts $(W, x^i, y^\mu)$ in the bundle $\pi$. Here $(\pi(W),x^i )$ is a chart in $X$ and $y^\mu$ are coordinates along the fibers.  Tangent frame corresponding to the local chart $(W, x^i, y^\mu)$ will be denoted by $(\partial_{i}=\partial_{x^i},\partial_{\mu}=\partial_{y^\mu})$ (shorter notation will be used in more cumbersome calculations), corresponding coframe - $(dx^i ,dy^\mu )$.\par  Introduce the contracted forms $\eta_{i}=i_{\partial_{x^i}}\eta,\ \eta_{ij}=i_{\partial_{x^j}}i_{\partial_{x^i}}\eta$. Below we will be using following relations for the forms  $\eta_{j},\eta_{ji}$ (here and below $\lambda_{G}=ln(\sqrt{\vert G\vert})$):
\beq \begin{cases}dx^j \wedge \eta_{i}=\delta^{j}_{i}\eta,\\  dx^j \wedge \eta_{ik }=\delta^{j}_{\sigma}\eta_{i}-\delta^{j}_{i}\eta_{k},\\ d\eta_{i}=\lambda_{G,i}\eta,\\ d\eta_{ij}=(\lambda_{G,j}\eta_{i}-\lambda_{G,i}\eta_{j})\end{cases}.\eeq \par

Sections $s:V\rightarrow Y,\ V\subset X$ of the bundle $\pi$ represent the collection of (classical) fields $y^\mu$ defined in the domain $V\subset X$. Usually these fields are components of some tensor fields or tensor densities fields (sections of "natural bundles", \cite{FF}).\par
    For a manifold $M$ we will denote by $\tau_{M}:T(M)\rightarrow M$ the tangent bundle of a manifold $M$, by $V(M)\subset T(M)$ the subbundle of of vertical vectors, i.e. vectors $\xi\in T_{y}(M)$ such that $\tau_{M*y}\xi =0$.\par

Denote by $\Lambda^{r}(M)$ the bundle of exterior $r$-forms on the manifold $M$ and by $(\Lambda^{*}=\oplus \Lambda^{r}(M),d)$ the differential algebra of exterior forms on the manifold $M$.\par
\subsection{The k-jet bundles $J^{k}(\pi)$.}

Given a fiber bundle $\pi:Y\rightarrow X$ denote by $J^{k}(\pi)$ the k-jet bundle of sections of the bundle $\pi$, \cite{GMS,KMS}.  Denote by $\pi_{kr}:J^{k}(\pi)\rightarrow J^{r}(\pi), k\geqq r\geqq 0$ the natural projections between the jet bundles of different order and by $\pi^{k}:J^{k}\rightarrow X$ the projection to the base manifold $X$.  Projection mappings $\pi_{k(k-1)}: J^{k}(\pi)\rightarrow J^{k-1}(\pi)$ in the tower of k-jet bundles  \[\ldots \rightarrow J^{k}(\pi)\rightarrow J^{k-1}(\pi)\rightarrow \ldots \rightarrow  Y \rightarrow X\]
are \emph{affine bundles} modeled by the vector bundle  $\bigwedge^{k} T^{*}(X)\otimes_{\rightarrow J^{k-1}(\pi)} V(\pi)\rightarrow J^{k-1}(\pi)$. \par Denote by $J^{\infty}(\pi)$ the infinite jet bundle of bundle $\pi$ - inverse limit of the projective sequence $\pi_{k(k-1)}:J^{k}(\pi)\rightarrow J^{k-1}(\pi)$. Space $J^{\infty}(\pi)$ is endowed with the structure of inverse limit of differentiable manifolds with the natural sheaves of vector fields, differential forms etc. making the projections $\pi_{\infty k}:J^{\infty}(\pi)\rightarrow J^{k}(\pi)$ smooth surjections.  See \cite{KMS,S} for more about structure and properties of k-jet bundles.\par

For a mutliindex $I =\{i_{1},\ldots ,i_{n}\},i_{k}\in \mathbb{N}$ denote by $\partial^{I}$ the differential operator in $C^{\infty}(X)$  $\partial^{I}f=\partial^{i_{1}}_{x^1}\cdot \ldots \partial^{i_{n}}_{x^n}.$  To every fibred chart $(V, x^i ,y^\mu )$ in $Y$
there corresponds the fibred chart $(x^i ,y^\mu ,z^{\mu}_{i},\vert I \vert =\sum_{s}i_{s}\leqq k)$ in the domain $V^k=\pi_{k0}^{-1}(V)\subset J^{k}(\pi)$.  This chart is defined by the condition $ z^{\mu}_{I}(j^{k}_{x}s)=\partial^I s^{\mu}(x).$\par

For $k=1,\ldots,\infty$ the space  of k-jet bundle $J^{k}(\pi)\rightarrow X$ is endowed with the \textbf{Cartan distribution} $Ca^{k}$ defined by the basic \textbf{contact forms}
\beq
\omega^{\mu}=dy^\mu -z^{\mu}_{i}dx^i,\ldots \omega^{\mu}_{I}=dz^{\mu}_{I}-z^{\mu}_{I +1_{j}}dx^j,\ \vert I\vert <k;\ dz^{\mu}_{I},\ \vert I\vert =k.
\eeq
 These forms generate the contact ideal $C^k  \subset \Lambda^{*}(J^{k}(\pi))$ in the algebra of all exterior forms. Denote by $I(C^k)$ the differential ideal of  contact forms.  This ideal is generated by the basic contact forms  (2.2) and by the forms $dz^{\mu}_{I},\ \vert I\vert =k$, see \cite{K3}. In the case where $k=\infty$ the basic contact forms $dz^{\mu}_{I}$ are absent form the list of generators of ideal $I(C^k)$.
 \par
 A $p$-form is called $l$-contact if it belongs to the l-th degree of this ideal $(C^k)^l \subset \Lambda^{*}(J^{k}(\pi)).$  0-contact forms are also called \textbf{horizontal (or $\pi_{k}$-horizontal ) forms} (or, sometimes, \emph{semi-basic forms}).  We denote by $kCon$ the $k$-contact forms that appear in calculations.  For $k=1$ we will omit index 1.\par

Let $0\leqq s<k$.  A form $\nu \in \Lambda^{*}(J^{k}(\pi))$ is called $\mathbf{\pi_{ks}}$-\textbf{horizontal} if it belongs to the subalgebra $C^{\infty}(J^{k}(\pi))\pi^{*}_{ks}\Lambda^{*}(J^{s}(\pi))\subset \Lambda^{*}(J^{k}(\pi)).$   \par

Remind now the following basic result (D. Krupka,\cite{K1})
\begin{theorem} Let a form $\nu \in \Lambda^{q}(J^{k}(\pi))$ be $\pi_{k(k-1)}$-horizontal (i.e. $\nu =\pi_{k(k-1)}\nu_{*},\nu_{*}\in \Lambda^{q}(J^{k-1}(\pi))$. Then there is unique \emph{contact} decomposition of the form $\nu$
\beq
\nu =\nu_{0}+\nu_{1}+\ldots +\nu_{q},
\eeq
where $\nu_{i},0\leqq i\leqq q$  is a $i$-contact form on $J^{k}(\pi).$  Form $\nu_{0}$ is called the horizontal part of the form $\nu_{*}$ (and of the form $\nu$ as well).
\end{theorem}
Related to the contact decomposition is the decomposition of differential operator $d$ as the sum of \textbf{horizontal and vertical differentials} $d_h ,d_v$: for a $q$-form $\nu \in \Lambda^{*}(J^{k})(\pi))$ its differential $d\nu$ lifted into the $J^{k+1}(\pi)$ is presented as the sum of horizontal and contact (vertical) terms:
\[
d\nu =d_{h}\nu+d_{v}\nu.
\]
,see~\cite{GMS,KV}. Operators $d_h ,d_v$ are more naturally defined in the space  $\Lambda^{*}(J^{\infty})(\pi))$.\par

We recall that these operators have the following homology properties
\[
d^{2}_{h}=d^{2}_{V}=d_{V}d_{h}+d_{h}d_{V}=0.
\]
\par
In particular, for a function $f\in C^{\infty}(J^{\infty}(\pi))$ (depending on the jet variables $z^{\mu}_{I}$ up to some degree, say $\vert I\vert \leqq k$),
\beq
df=(d_{i}f)dx^i +\sum_{\vert I\vert \geqq 0}f_{,z^{\mu}_{I}}\omega^{\mu}_{I },
\eeq
where
\beq
d_{i}f =\partial_{x^i}f+\sum_{I\vert \vert I\vert \geqq 0} z^{\mu}_{I +1_{i}}\partial_{z^{\mu}_{I}}f
\eeq
is the \textbf{total derivative} of the function $f$ by $x^i$.  The series in the formulas (2.4-5) contains finite number of terms: $\vert I\vert \leqq k$.\par

\subsection{Lagrangian Poincare-Cartan formalism, k=1.}  Here we remind the basic notions of infinitesimal form of Lagrangian Field Theory of the first order (on $J^{1}(\pi)$) based on the use of Poincare-Cartan form,\cite{FF,GMS,LMD}.\par
Volume form $\eta $ permits to construct the
{\bf vertical endomorphism}
\begin{equation}
S_{\eta }=(dy^{\mu}-z^{\mu}_{i }dx^{i} )\wedge \eta_{j}\otimes \frac{\partial }{\partial z^{\mu}_{j}}
\end{equation}
which is a tensor field of type $(1, n)$ on the 1-jet bundle $J^{1}(\pi)$.\par
For a Lagrangian n-form $L\eta$, $L$ being a (smooth) function on the manifold $J^{1}(\pi)$ the
Poincar\'e-Cartan $n$-form are defined as follows:
\begin{equation}
\Theta _{L}=L\eta +S^{*}_{\eta }(dL),
\end{equation}
where $S^{*}_{\eta }$ is the adjoint operator of $S_{\eta}$. In coordinates we have
\begin{equation}
\Theta _{L} = (L-z^{\mu}_{i} \frac{\partial L}{\partial z^{\mu}_{i}})\eta + \frac{\partial L}{\partial
z^{\mu}_{i}}dy^{\mu}\wedge \eta_{i},
\end{equation}
An {\it extremal} of $L$ is a section $s$ of the bundle $\pi$ such that for any vector field $\hat{\xi}$ on the manifold $J^{1}(\pi)$,
\begin{equation}
(j^{1}(s) )^{*}(i_{\hat{\xi} }d\Theta_{L})=0,
\end{equation}
where $j^{1}(s) $ is the first jet prolongation of $s $.  A section $s$ is an extremal of $L$ if and only if it satisfies to the system of Euler-Lagrange Equations (see, for instance, \cite{BSF,GMS})
\beq  \frac{\partial (L\circ j^{1}(s) \sqrt{\vert G\vert})}{\partial y^{\mu}}-\frac{d }{d
x^{i}}\left( \frac{\partial (L\circ j^{1}(s) \sqrt{\vert G\vert})}{\partial z^{\mu}_{i}}\right)  =0,\ 1\leqq
\mu\leqq m. \eeq

\subsection{Bundles $\Lambda^{r}_{p}Y$ and canonical forms.}

Introduce the bundle $\Lambda^{r}_{p}(Y)$ of the exterior forms on $Y$ which are annulated if p of
its arguments are vertical (\cite{K2,LMD}:
\[ \omega^{r}\in  \Lambda^{r}_{p}(Y) \Leftrightarrow  i_{\xi_{1}} \ldots  i_{\xi_{p}}\omega^{r}=0,\ \xi_{i}\in V(Y).\]
We will be using these bundles for $r=n,n+1,n+2$ and $p=1,2$.\par
 Elements of the space $\Lambda ^{n}_{1}Y$ are semibasic $n$-forms locally expressed as $p(x,y) \eta
 .$ \par
 Elements of the space $\Lambda^{n}_{2}Y$ have, in a fiber coordinates $(x^i, y^\mu )$ the form
\[ p(x,y) \eta + p_{\mu}^{i}dy^{\mu}\wedge \eta_{i}.\]\par
 This introduces coordinates $(x^{i},y^{\mu},p)$ on the manifold $\Lambda ^{n}_{1}Y$ and $(x^{i},y^{\mu},p,p_{\mu}^{i})$ on the manifold $\Lambda^{n}_{2}Y.$\par

For the case where $r=n+1$ the forms $dy^\mu\wedge \eta$ form the basis of fibers of the bundle $\Lambda^{n+1}_{2}(Y)\rightarrow Y$
while the bundle $\Lambda^{n+1}_{1}(Y)$ is \emph{zero bundle}.\par
Introduce the notation
\[\Lambda^{n+(n+1)}_{p}(Y)=\Lambda^{n}_{p}(Y)\oplus
\Lambda^{n+1}_{p}(Y),p=1,2,\]
for the direct sum of the bundles on the right side.\par
It is clear that $\Lambda^{k}_{1}(Y)\subset \Lambda^{k}_{2}(Y)$. Therefore we have the embedding of subbundles $\Lambda^{n+(n+1)}_{1}(Y)\subset \Lambda^{n+(n+1)}_{2}(Y)$
and can define the factor-bundle \[ \Lambda^{n+(n+1)}_{2/1}(Y)=\Lambda^{n+(n+1)}_{2}(Y)/\Lambda^{n+(n+1)}_{1}(Y), \]
with the projection $q:\Lambda^{n+(n+1)}_{2}(Y)\rightarrow \Lambda^{n+(n+1)}_{2/1}(Y) $.\par
On the bundles $\Lambda^{n+(n+1)}_{2}(Y)$ there is defined the canonical form (\cite{LMD}) with the coordinate expression
\beq \Theta^{n+(n+1)}_{2}=\Theta^{n}_{2}+\Theta^{n+1}_{2}= p\eta+p_{\mu}^{i}dy^{\mu}\wedge \eta_{i}+p_{\mu}dy^\mu \wedge \eta.\eeq
On the factor-bundle $\Lambda^{n+(n+1)}_{2/1}(Y)$ the form \[\widetilde{\Theta}^{n+(n+1)}_{2/1}=p_{\mu}^{i}dy^{\mu}\wedge \eta_{i}+p_{\mu}dy^\mu \wedge \eta\]
is defined $mod\ \eta$.\par
\subsection{Legendre transformations,\ $k=1$.}
Let $L\in C^{\infty}(J^{1}(\pi))$ be a Lagrangian function. We define the fiber mapping (over $Y$) $leg_{L}:J^{1}(\pi)\rightarrow \Lambda^{n}_{2}Y,$
as follows:
\[
leg_{L}(j^{1}_{x}s))(X_{1},\ldots, X_{n}) =(\Theta_{L})_{j^{1}_{x}s}({\tilde{X}_{1},\ldots,
\tilde{X}_{n}}),
\]
where $j^{1}_{x}s \in J^{1}(\pi), X_{i}\in T_{s (x)}Y$ and $\tilde{X}_{i}\in T_{j^{1}_{x}s (X)}J^{1}(\pi)$ are such
that $\pi_{*}(\tilde{X}_{i})=X_{i}.$  In local coordinates, we have
\[
leg_{L}(x^{i},y^{\mu},z^{\mu}_{i}) = (x^{i},y^{\mu},p=L-z^{\mu}_{i}\frac{\partial L }{\partial
z^{\mu}_{i}},p^{i}_{\mu} =\frac{\partial L }{\partial z^{\mu}_{i}}).
\]
The Legendre transformation $Leg_{L}:J^{1}(\pi)\rightarrow \Lambda^{n+(n+1)}_{2/1}(Y)$ is defined as the composition $Leg_{L}=q\circ leg_{L}.$ In coordinates,
\[
Leg_{L}(x^{i},y^{\mu},z^{\mu}_{i})=(x^{i},y^{\mu},p^{i}_{\mu}=\frac{\partial L }{\partial z^{\mu}_{i}}).
\]
Recall \cite{LMD} that the Legendre transformation $Leg_{L}:J^{1}(\pi)\rightarrow \Lambda^{n+(n+1)}_{2/1}(Y)$ is a local diffeomorphism
if and only if $L$ is regular, i.e. when the\emph{ vertical Hessian} $\frac{\partial^{2}L}{\partial z^{\mu}_{i} \partial z^{\nu}_{j}}$ is \emph{nondegenerate}. \par
\subsection{Con-differential.}
Below we will be using the con-differential ${\tilde d}$ used in topology, see. for instance, \cite{IW}.\par

Let $M^n$ be an n-dimensional smooth manifold.  Con-differential is defined on couples of exterior forms $\alpha^{k}+\beta^{k+1}\in \Lambda^{k+(k+1)}(M)$:
\beq\begin{cases}
 {\tilde d}: \Lambda^{k+(k+1)}(M)=\Lambda^{k}(M)\oplus
\Lambda^{k+1}(M)\rightarrow \Lambda^{(k+1)+(k+2)}(M)=\Lambda^{k+1}(M)\oplus \Lambda^{k+2}(M):\\ {\tilde
d}(\alpha^{k}+\beta^{k+1} )=((-d\alpha+\beta )+d\beta ).\end{cases} \eeq
\begin{lemma} ${\tilde d}\circ {\tilde d}=0.$
\end{lemma}
\begin{proof} We have
\[{\tilde d} {\tilde d} (\alpha^{k}+\beta^{k+1}) = {\tilde d} ((-d\alpha+\beta )+d\beta )=
[-d(-d\alpha+\beta )+d\beta ]+d(d\beta)=-d\beta +d\beta +0.
\] \end{proof}
Some other properties of this differential and of the corresponding complex $(\Lambda^{k+(k+1)}(M)=\Lambda^{k}(M)\oplus \Lambda^{k+1}(M),\tilde d)$ are presented in \cite{Pr1}.
\subsection{Balance systems}
Let the base manifold $X$ be the material or physical space-time $X$ with (local) coordinates $x^1=t,\ x^A, A=1,2,3.$
Typical system of balance equations for the fields $y^\mu$ is determined by a choice of flux fields $F^{i}_{\mu},\ \mu=0,1,2,3$ (including the densities for $i=0$) and source fields $\Pi_{\mu}$ as functions on a jet space $J^{k}(\pi)$. Choice of the functions $F^{i}_{\mu},\Pi_{\mu}\in C^{\infty}(J^{k}(\pi))$ is codified in physics as the choice of the \textbf{constitutive relation} of a given material, \cite{Mu,MPE}. After such a choice has been done, the \emph{closed} system of balance equations for the fields $y^i$
\beq
( F^{i}_{\mu}\circ j^{k}s )_{;\mu}=\Pi_{\mu}\circ j^{k}s,\ \mu =1,\ldots, m
\eeq
can be solved for a section $s:V\rightarrow Y$ if one add to the balance system the appropriate boundary (including initial) conditions.\par
Introducing the horizontal forms in $J^{k}(\pi)$ -  $F_{\mu}=F^{i}_{\mu}\eta_{i},\Pi_{\mu}=\Pi_{\mu}\eta$ these equations can be written in the form \[j^{1\ *}(s)[dF_{\mu}-\Pi_{\mu}]=0.\]\par
Using the con-differential $\tilde d$ (see above) the balance system (2.13) can be written in the compact form
\beq j^{1\ *}(s){\tilde d}[F_{\mu}+\Pi_{\mu}]=0,\ i=1,\ldots, m.
\eeq
\begin{remark} Similarly to the definition of a conservation law on a contact manifold given in \cite{BGG} one can define a balance law of order $k$ as a $n+(n+1)$-form $\sigma =F^{n}+\Pi^{n+1}$ on the space $J^{k}(\pi)$ (which usually is $\pi_{k}$-horizontal) where $F^n$ is defined mod closed forms from  $Z^{n}(J^{k}(\pi))$ such that
\beq
\tilde d \sigma =dF^{n}-\Pi^{n+1}\in I(C^k ).
\eeq
In Continuum Thermodynamics balance laws typically are present as the closed system of equations for the dynamical fields.  It is shown below that the Poincare-Cartan formalism allows to generate the system of balance laws in the number equal to the number of dynamical fields $y^\mu$.
\end{remark}
\begin{example}{\textbf{Five fields fluid thermodynamical system.}}  In the \emph{5 fields fluid} thermodynamical system (\cite{Mu,MPE}), the velocity vector field $\mathbf{v}$, density scalar field $\rho$ and scalar field of internal energy $\epsilon$ (or total energy $e =\frac{1}{2}\rho\Vert v\Vert^{2}+\epsilon $) are considered as the basic dynamical fields $y^\mu$. Corresponding balance system has the form of the five balance equations - mass conservation law, linear momentum balance law and the energy balance law:
\beq
\begin{cases}
\partial_{t}\rho +\partial_{x^B}(\rho v^B )=0,\\
\partial_{t}(\rho v^A )+\partial_{x^B}(\frac{1}{2}\rho v^{A}v^{B}+t^{AB})=f^{A},\ A=1,2,3,\\
\partial_{t}(\frac{1}{2}\rho\Vert v\Vert^{2}+e)+
\partial_{x^B}\left( (\frac{1}{2}\rho \Vert v\Vert^{2}+e) v^B+t_{C}^{B}v^{C}-h^B \right)=f_{B}v^B +r.
\end{cases}
\eeq
Here $\mathbf{t}$ is the (1,1)-stress tensor, $\mathbf{f}$ is the 1-form of the bulk force, $\mathbf{h}$ is the heat flux vector filed and the scalar function  $r$ - radiation heating source.\par
\end{example}
\begin{example}
\end{example}
\section{Partial k-jet bundles}
In this section we introduce the "partial jet bundles" of the configurational bundle $\pi:Y\rightarrow X$ - factor bundles of the conventional jet bundles $J^{k}(\pi)$. These bundles are convenient for the description of balance systems whose constitutive relations depends on some but not all derivatives of fields $y^\mu$.  We start with the definition of the partial 1-jet bundle defined by a subbundle $K\subset T(X)$. \par
\subsection{Bundle $J^{1}_{K}(\pi)$.}
\begin{definition} Let $\pi_{XY}:Y\rightarrow X$ be a fiber bundle and let $K\subset T(X)$ be a subbundle. \begin{enumerate}
\item Let $x\in U\subset X,$ $s_{1},s_{2}\in \Gamma (U,\pi)\vert s_{1}(x)=s_{2}(x)$. Sections $s_{1},s_{2}$ are called $K$-\textbf{equivalent of order 1 at a point} $x\in X:\ s_{1}\sim_{K_{x}} s_{2} $ if
$  s_{1*x}\vert_{K_{x}}=s_{2*x}\vert_{K_{x}}.$
\item  $J^{1}_{K\ x}(\pi)$- space of classes of $\sim_{K_{x}}$  at a point $x\in X$.
\item $J^{1}_{K}(\pi)=\cup_{x\in X}J^{1}_{K\ x}(\pi)$ - the space of partial 1-jets of sections of $\pi$.
\end{enumerate}
\end{definition}
In the next proposition we collect basic properties of bundles $J^{1}_{K}(\pi)$. Proof of this Proposition is straightforward ( see \cite{Pr1}).
\begin{proposition}
\begin{enumerate}
\item  Bundle $J^{1}_{K}(\pi)\rightarrow Y$ is the affine bundle modeled on the vector bundle
$\pi^{*}(K^{*})\otimes V(\pi)\rightarrow Y$. \par
\item  There is a canonical surjection of affine bundles $ w_{K}: J^{1}(\pi)\rightarrow  J^{1}_{K}(\pi).$
\item Let $T(X)=K(X)\oplus K'(X)$ be an almost product structure (AP), then there is the
commutative diagram
\[
\begin{CD}  J^{1}(\pi) @> w_{K} >> J^{1}_{K}(\pi)\\
@V w_{K'} VV        @V \pi_{10\ K} VV\\
 J^{1}_{K'}(\pi) @>\pi_{10\ K'} >>  Y,
\end{CD}
\]
which realizes the conventional 1-jet bundle $J^{1}(\pi)$ as the fiber product of the bundles $J^{1}_{K}(\pi),J^{1}_{K'}(\pi)$.
\item Let $T(X)=K\oplus K'$ be an integrable AP-structure, $(x^j,x^k )$ - local chart integrating the $AP-structure$ (i.e. $K_{x}$ is the linear span of vector fields $\partial_{j}$, $K'$ - is the linear span of vector fields $\partial_{k}$). One forms
\[ \omega^{\mu}=dy^\mu-\sum_{j }z^{\mu}_{j}dx^j \]
generate the "partial" Cartan distribution $C_{K}$ on $J^{1}_{K}(\pi )$ in the sense that a section $q$ of the bundle $J^{1}_{K}(\pi )$ is the (partial) 1-jet of a section $s:X\rightarrow Y$ if and only if $q^{*}(\omega^\mu )\vert_{K}=0,\ \forall \mu=1,\ldots ,m.$
\item All constructions above are covariant with respect to the automorphisms $\phi$ of the bundle $\pi$ such that projection $\bar \phi$ of automorphism $\phi$ to the base $X$ leaves the AP structure $T=K\oplus K'$ invariant.
\end{enumerate}
\end{proposition}
\begin{remark} Partial Cartan structure exist in the case of a general AP structure on the manifold $X$, see \cite{Pr1} for the proof and construction of the basic contact forms.
\end{remark}
\begin{example} Basic example is, of course, the space-time decomposition where $X=T\times B$ is the product of the classical time axis and the 3-dim space manifold $B$ (material or physical) and where $T(X)=\langle \partial_{t}\rangle \oplus T(B)$.  In such a case $x^1=t$ and $x^i,i=2,3,4$ are spacial coordinates.  We take $K=T(B)\subset T(X)$ to be the subbundle of derivatives in spacial directions.
\end{example}
\begin{example} Mathematically trivial but very important physically is the case of the Rational Extended Thermodynamics, \cite{MR}, where the configurational space is extended enough so that the constitutive relations do not depend on the derivatives of the basic fields $y^\mu$.  To include this case in our scheme we take $K=\{ 0\}.$ Then, the bundle $J^{1}_{\{0\}}(\pi )\rightarrow Y$ has zero dimensional fiber.
\end{example}
\subsection{Space-time splitting bundle $J^{1}_{S}(\pi)$.}
Employing the construction of the bundle $J^{1}_{K}(\pi)$ for the product structure $T(X)=\langle \partial_{t}\rangle \oplus T(B)$ we define now the partial 1-jet bundle $J^{1}_{S}(\pi).$ We assume that the generic fiber $U$ of the bundle $\pi:Y\rightarrow X$ is the direct product of subspaces of fields $y^i$ entering the constitutive relation without derivatives, with time  derivative only, with spacial derivatives only and with all derivatives respectively:\hfill
\[ [1,\ldots ,m]=S_{0}\cup S_{t}\cup S_{x}\cup S_{xt} \Rightarrow U=U_{0}\times U_{t}\times U_{x}\times U_{tx}. \]
\begin{proposition}
\begin{enumerate} \item Bundle $\pi:Y\rightarrow X$ is the fiber product of the  bundles $\pi_{0}:Y_{0}\rightarrow X,\ \pi_{t}:Y_{t}\rightarrow X,\ldots $ with fibers $U_{0},U_{t},\ldots $ over the base manifold $X:$
$ Y=Y_{0}\underset{X}{\times}  Y_{t}\underset{X}{\times}
Y_{x}\underset{X}{\times}  Y_{xt}.$
\item The bundle $J^{1}_{S}(\pi)$ is the fiber product of \textbf{affine bundles}
\[ J^{1}_{S}(\pi)= 0(Y_{0})\underset{Y}{\times} J^{1}_{\langle \partial_t \rangle
}(Y_{t})\underset{Y}{\times} J^{1}_{\langle \partial_{x^A }\rangle }(Y_{x})\underset{Y}{\times}
J^{1}(Y_{tx}). \]
 \item  Bundle $J^{1}_{S}(\pi)\rightarrow Y$ is the factor-bundle of the bundle $J^{1}(\pi)$.
\item Partial Cartan distribution $C_{S}$ is generated by the 1-forms
\[
\omega^{\mu}=dy^\mu -z^{\mu}_{t}dt,\ \mu\in S_{t},\ \omega^{\mu}=dy^\mu -z^{\mu}_{A}dx^A,\ \mu\in S_{x}; \omega^{\mu}=dy^\mu
-z^{\mu}_{i}dx^i,\ \mu\in S_{tx}
\]
in the sense that a section $\sigma:X\rightarrow J^{1}_{S}(\pi)$ is integrable: $\sigma =j^{1}_{p}(s)$ for a section $s:X\rightarrow Y$ if and only if $\ \text{for\ all}\ i.$ $\sigma^{*}\omega^\mu =0$ if restricted to the corresponding distribution.
 \end{enumerate}
\end{proposition}
\par
\begin{remark} Denote by  $Aut_{S}(\pi)$ the subgroup of the group $Aut(\pi)$ consisting of the transformations $\phi \in Aut_{S}(\pi)$ such that $\bar \phi$ preserves the space-time product structure and, in addition, $\phi$ is the fiber product of automorphisms of the bundles $\pi_{0},\pi_{t},\pi_{x}\pi_{tx}.$ Construction of the bundle $J^{1}_{S}(\pi)$ is invariant with respect to the transformations $\phi \in Aut_{S}(\pi)$.
\end{remark}

\subsection{Prolongation of vector fields to the partial 1-jet bundles.}

Below we will use the prolongations of the automorphisms $\phi$ of the bundle $\pi$ whose projection $\bar \phi$ to $X$ preserves the AP structure $T(X)=K\oplus K'$ (respectively,  an $S$-structure) to the partial 1-jet bundle $J^{1}_{K}(\pi)$ (respectively to $J^{1}_{S}(\pi)$).  Infinitesimal version of this procedure is the prolongation of a $\pi$-projectable vector field $\xi$ whose projection $\bar \xi$ generates a (local) phase flow preserving the subbundle $K$ (respectively, $S$-structure).  In the case of a conventional k-jet bundle $J^{k}(\pi)$ this procedure is well known (\cite{GMS,KV, O}).  In our situation the results are similar.  We formulate corresponding results in the infinitesimal case for the bundle $J^{1}_{K}(\pi)$ with an integrable AP structure.  For more general case of an arbitrary AP-structure we refer to \cite{Pr1}.\par
\begin{definition} \begin{enumerate}
\item Denote by $\mathcal{X}_{K}(\pi )$ the Lie algebra of  $\pi$-projectable vector fields
 $\xi $ in $Y$ such that the projection $\bar \xi$ of $\xi $ to $X$ \emph{preserves the distribution}
 $K\subset T(M)$: ${\bar \phi}_{t*}K =K$ for the local flow ${\bar \phi}_{t}$ of $\bar \xi .$  This condition is equivalent to the infinitesimal condition $\mathcal{L}_{\bar \xi}K\subset K$.
\item
 Denote by $\mathcal{X}_{K\oplus K'}(\pi )$ the Lie algebra of  $\pi$-projectable vector fields
 $\xi $ in $Y$ such that the field $\bar \xi$ generated by $\xi $ in $X$ \emph{preserves the AP-structure}, \cite{LR}
 $T(M)=K\oplus K'$: ${\bar \phi}_{t*}K =K,{\bar \phi}_{t*}K' =K'$ for the local flow ${\bar \phi}_{t}$ of $\bar \xi$.
 \end{enumerate}
\end{definition}
\begin{lemma} Let the AP-structure $T(X)=K\oplus K'$ is integrable and let $(x^j ,x^k )$ be  a
(local) integrating chart (i.e. $K= <\partial_{x^j}>;\ K'=<\partial_{x^k}>$). Then
\begin{enumerate}
\item A $\pi$-projectable vector field $\xi=\xi^{i}(x)\partial_{x^i}+\xi^{\mu}(x,y)\partial_{y^\mu}$ belongs
to $\mathcal{X}_{K}(\pi )$ if and only if
\[
{\bar \xi}=\xi^{i}(x^j ,x^k )\partial_{x^i}=\xi^{j}(x)\partial_{x^j}+\xi^{k}(x^k
)\partial_{x^k},
\]
i.e. if the components $\xi^{k}(x)$ do not depend on the variables $x^j$.
\item A $\pi$-projectable vector field $\xi=\xi^{i}(x)\partial_{x^i}+\xi^{\mu}(x,y)\partial_{y^\mu}$
belongs to $\mathcal{X}_{K\oplus K'}(\pi )$ (preserves the almost product structure $T(X)=K\oplus K'$) if
and only if
\[
{\bar \xi}=\xi^{i}(x )\partial_{x^i}=\xi^{j}(x^{j_{1}})\partial_{x^j}+\xi^{k}(x^k
)\partial_{x^k},
\]
\end{enumerate}
\end{lemma}
\begin{proof}  We have
$
[{\bar \xi},\partial_{x^j}]=-(\partial_{x^j}\cdot
\xi^{j_{1}})\partial_{x^{j_{1}}}-(\partial_{x^j}\cdot \xi^{k})\partial_{x^k}.
$ This vector field belongs to $K$ if and only if $\partial_{x^j}\cdot \xi^{k}=0$ for all $j$ and
$k$.  The second statement is proved in the same way.
\end{proof}
\begin{proposition} Let the AP-structure $T(X)=K\oplus K'$ is integrable and let $(x^j ,x^k )$ be a
(local) integrating chart . \begin{enumerate}\item For a $\pi$-projectable vector field
$\xi=\xi^{i}(x)\partial_{x^i}+\xi^{\mu}(x,y)\partial_{y^\mu} \in \mathcal{X}_{K}(\pi)$ the following
properties are equivalent
\begin{enumerate}
\item There exist a vector field $\xi^1 \in \mathcal{X}(J^{1}_{K}(\pi))$ such that
\begin{enumerate}
\item Local flow of the vector field $\xi^1$ preserves the partial Cartan distribution $Ca_{K}$ ($\xi^1$ is the {\emph{Lie field}} in terminology of  \cite{KV}).
\item $\pi_{10\ *}\xi^1 =\xi$.
\end{enumerate}
\item Vector field $\xi$ has, in a local integrating chart $(x^j ,x^k)$ the form
\beq
\xi
=\xi^{j}(x^{j_{1}})\partial_{x^{j}}+\xi^{k}(x^{k_{1}})\partial_{x^{k}}+\xi^{\mu}(x^j
,y)\partial_{y^\mu}.
\eeq
In particular the projection $\bar \xi$ of the vector field $\xi $ in $X$ \textbf{preserves the almost product
structure} $K\oplus K'$.
\item Vice versa, any Lie vector field on $J^{1}_{K}(\pi)$ is the prolongation of a vector field $\xi \in \mathcal{X}_{K\oplus K'}(Y)$ of the form (3.1) (see \cite{KV} Ch.2).
\end{enumerate}
\item In the case where these conditions are fulfilled the vector field $\xi^1$ is unique and has the form
\beq
\xi^1=\xi^{j}(x^{j_{1}})\partial_{x^{j}}+\xi^{k}(x^{k_{1}})\partial_{x^{k}}+\xi^{\mu}(x^j
,y)\partial_{y^\mu}+\left(d_{j}\xi^\mu -z^{\mu}_{j_{1}}\frac{\partial \xi^{j_{1}}}{\partial x^j} \right)
\partial_{z^{\mu}_{j}}\eeq
\item Mapping $\xi \rightarrow \xi^{1}$ is the homomorphism of Lie algebras:
$ [\xi,\eta]^{1}=[\xi^{1},\eta^{1}] $ for all $\xi,\eta \in \mathcal{X}_{K,K'}(\pi).$
\end{enumerate}
\end{proposition}
\begin{proof} Let
${\hat
\xi}=\xi^{i}(x)\partial_{x^i}+\xi^{\mu}(x,y)\partial_{y^\mu}+\lambda^{\mu}_{j}\partial_{z^{\mu}_{j}}$ be a
prolongation to the partial jet bundle $J^{1}_{K}(\pi)$ of the vector field $\xi$.  Then, condition of the
preservation of the partial Cartan structure is equivalent to the condition that for all the generators
$\omega^{\mu}_{K}=dy^\mu-\sum_{j}z^{\mu}_{j}dx^j$ of the contact ideal of exterior forms,
$\mathcal{L}_{\hat \xi}\omega^{\mu}_{K}=\sum_{\nu}q^{\mu}_{\nu}\omega^{\nu}_{K},$ for some functions $q^{\mu}_{\nu}\in C^{\infty}(J^{1}_{K}(\pi))$. We calculate
\begin{multline}
\mathcal{L}_{\hat \xi}\omega^{\mu}_{K}=(di_{\hat \xi}+i_{\hat \xi}d)(dy^\mu-\sum_{j}z^{\mu}_{j}dx^j)=
d[\xi^\mu -z^{\mu}_{j}\xi^{j}]+i_{\hat \xi }(-dz^{\mu}_{j}\wedge dx^j )=\\
=d\xi^{\mu}-\xi^{j}dz^{\mu}_{j}-z^{\mu}_{j}d\xi^{j}-\lambda^{\mu}_{j}dx^j+\xi^{j}dz^{\mu}_{j} =
\xi^{\mu}_{,x^i}dx^i+\xi^{\mu}_{,y^\nu}dy^\nu-z^{\mu}_{j}[\xi^{j}_{,x^{j_{1}}}dx^{j_{1}}
+\xi^{j}_{,x^k}dxk]-\\
-\lambda^{\mu}_{j}dx^j =\sum_{\nu}q^{\mu}_{\nu}(dy^\nu-\sum_{j}z^{\nu}_{j}dx^j),
\end{multline}
or
\[
(\xi^{\mu}_{,x^k}-z^{\mu}_{j}\xi^{j}_{,x^k})dx^k +\xi^{i}_{,y^j}dy^j+[\xi^{i}_{,x^j}
-\lambda^{\mu}_{j}-z^{\mu}_{j_{1}}\xi^{j_{1}}_{,x^{j}}]dx^j =
\sum_{\nu}q^{\mu}_{\nu}(dy^\nu-\sum_{j}z^{\nu}_{j}dx^j).
\]
This equality is fulfilled if and only if we have
\[
\begin{cases}
\xi^{\mu}_{,x^k}-z^{\mu}_{j}\xi^{j}_{,x^k}=0,\\
q^{\mu}_{\nu}=\xi^{\mu}_{,y^\nu},\\
\xi^{\mu}_{,x^j } -\lambda^{\mu}_{j}-z^{\mu}_{j_{1}}\xi^{j_{1}}_{,x^{j}}=-q^{\mu}_{\nu}z^{\nu}_{j}.
\end{cases}
\]
Since neither $\xi^\mu$ nor $\xi^j$ depend on $z^{\mu}_{i}$ first system is equivalent to the requirement
that both $\xi^\mu$ and $\xi^j$ are independent on $x^k$. Then the second condition determines
$q^{\mu}_{\nu}$ and third - $\lambda^{\mu}_{j}=\xi^{\mu}_{,x^j
}+\xi^{\mu}_{,y^\nu}z^{\nu}_{j}-z^{i}_{j_{1}}\xi^{j_{1}}_{,x^{j}}$ and the prolongation $\hat \xi$ takes
the form described in the Proposition.
\end{proof}
Similar results have place for the bundles $J^{1}_{S}(\pi)$ defined by a space-time splitting $S$ of the fields $y^\mu$.
\begin{proposition}
\begin{enumerate}
\item A vector field $\xi\in \mathcal{X}_{S}(\pi)$ preserves the AP-structure $T(X)=T(B)\oplus <\partial_{t}>$
if and only if ${\bar \xi}=\xi^{i}(x,t)\partial_{x^i}=\xi^A (x)\partial_{x^A}+\xi^{t}(t )\partial_{t},$
\item
For any $S$-admissible $\pi$-projectable vector field $\xi \in \mathcal{X}_{S}(\pi)$ following statements
are equivalent
\begin{enumerate}
 \item There is a vector field $\xi^{1}\in \mathcal{X}(J^{1}_{S}(\pi))$  such that
\begin{enumerate}
\item Vector field $\xi^{1} \in \mathcal{X}(J^{1}_{S}(\pi))$ is $\pi_{10}$-projectable and $\pi_{10*}(\xi^1 )=\xi,$
\item Local flow of the vector field $\xi^{1}$ preserves the partial
Cartan distribution $Co_{S}$ at $J^{1}_{S}(\pi)$.
\end{enumerate}
\end{enumerate}
\item Mapping $\xi \rightarrow \xi^{1}$ is the homomorphism of Lie algebras:
$[\xi,\eta]^{1}=[\xi^{1},\eta^{1}] $ for all $\xi,\eta \in \mathcal{X}_{K}(Y).$
\end{enumerate}

\end{proposition}

Detailed structure of vector fields $\xi$ preserving the $S$-structure and form of their prolongations $\xi^1$ to the bundle $J^{1}_{S}(\pi)$ is described in \cite{Pr1}, Ch.2, Prop. 11.
\subsubsection{Prolongation of infintesimal automorphisms $\xi \in \mathcal{X}_{p}(\pi)$ to the bundles $\Lambda^{k}_{r}(Y)$, $\Lambda^{n+(n+1)}_{2/1}(J^{1}(\pi))$.}
\begin{definition}(Definition-Proposition, \cite{LDS}, Def.3.3. for a case of euclidian metric $G$) Let $\alpha$ be a pullback to
$\Lambda^{n}_{2}Y$ of a $\pi$-semibasic form $\alpha =\alpha^{j}(x,y)\eta_\nu$ on $Y$. Let $\xi \in
\mathcal{X}(Y)$.
\begin{enumerate}
\item
 Then there exist and is unique a vector field
$\xi^{*\alpha}$ on $\Lambda^{n}_{2}$ satisfying to the following conditions
\begin{enumerate}
\item Vector field $\xi^{*\alpha}$ is $\pi_{\Lambda^{n}_{2}Y  \ Y}$-projectable and
\[
\pi_{\Lambda^{n}_{2}Y  \ Y\ *} \xi^{*\alpha}=\xi,
\]
\item
\[
\mathcal{L}_{\xi^{*\alpha}}\Theta^{n}_{2}=d\alpha.
\]
\end{enumerate}
\item In an adopted chart $(x^\mu ,y^i )$ the vector field $\xi^{*\alpha}$ has the form
\begin{multline}
\xi^{*\alpha}=\xi +\xi^{*\alpha \ p}\partial_{p}+\xi^{*\alpha \ p^{i}_{\mu}}\partial_{p^{i}_{\mu}},\ \text{where} \\
\xi^{*\alpha \ p}= -p\left( \frac{\partial \xi^i }{\partial x^i }+\xi^{i}\lambda_{G,i}\right) -p^{i}_{\mu}\frac{\partial \xi^{\mu}}{\partial x^i }+\left( \frac{\partial \alpha^i }{\partial x^i }-\alpha^{j}\lambda_{G,j}\right) =-p\cdot div_{G}({\bar \xi})-p^{\mu}_{\mu}\frac{\partial \xi^{\mu}}{\partial x^i }+div_{G}({\bar \alpha});\\
\xi^{*\alpha \ p^{i}_{\mu}}= p^{j}_{\mu}\frac{\partial \xi^i }{\partial x^j
}-p^{i}_{\nu}\frac{\partial \xi^\nu }{\partial y^\mu}-p^{i}_{\mu}\left(
\frac{\partial \xi^j }{\partial x^j
}+\xi^{j}\lambda_{G,j}\right)+\frac{\partial \alpha^i
}{\partial y^\mu}=  p^{j}_{\mu}\frac{\partial \xi^i}{\partial x^j
}-p^{i}_{\nu}\frac{\partial \xi^\nu }{\partial y^\mu}-p^{i}_{\mu}\cdot  div_{G}({\bar \xi})+\frac{\partial \alpha^i
}{\partial y^\mu}.
\end{multline}
Here ${\bar \xi}=\xi^{i}\partial_{x^i},\ \overline{\alpha }=\alpha^{j}(x,y)\partial_{x^j}.$
\item Let a vector field $\xi \in \mathcal{X}(Y)$ be $\pi$-projectable.  Then the 0-lift $\xi^{*0}$  of $\xi $
coincide with the flow prolongation $\xi^{ 1*}$ defined above.
\end{enumerate}
\end{definition}
Below we will use the lift of a projectable vector field $\xi$ to the bundle $\Lambda^{(n)+(n+1)}_{2/1}Y$.  Next result  allows to lift $\xi$ to the infinitesimal transformation on the bundle $\Lambda^{(n+1)}_{2}Y$ leaving invariant the canonical form  $p_{\mu}\omega^\mu \wedge \eta$.

\begin{proposition} For any $\pi$-projectable vector field  $\xi \in X(\pi)$ there exists unique projectable (to $Y$)
vector field $\xi^{*(n+1)}$ on the bundle $\Lambda^{(n+1)}_{2}Y$ that leaves the canonical form
$p_{\sigma }dy^{\sigma}\wedge \eta$ invariant.  That vector is given by the relation

\beq \xi^{*(n+1)}=\xi + \xi^{p_\sigma }\partial_{p_{\sigma}},\ \xi^{p_\sigma}= -p_{\sigma}(\xi^{i}\lambda_{G,i}
+\frac{\partial \xi^i}{\partial x^i})-p_{\nu}\frac{\partial \xi^\nu}{\partial y^\sigma}=-p_{\sigma} div_{G}({\bar \xi})-p_{\nu}\frac{\partial \xi^\nu}{\partial y^\sigma}, \eeq

where $\lambda_{G} =ln(\vert G\vert)$.
\end{proposition}
\begin{proof} We have, for a vector field of the form  $\xi^{*(n+2)}=\xi + \xi^{p_\sigma}\partial_{p_\sigma}$
\begin{multline}
\mathcal{L}_{\xi^{*(n+1)}}(p_{\mu}dy^{\mu}\wedge \eta)=\\=
(di_{\xi^{*(n+1)}}+i_{\xi^{*(n+1)}}d)(p_{\mu}dy^{\mu}\wedge \eta)=d[p_{\mu}\xi^{\mu}\eta-p_{\mu}\xi^{i}dy^\mu \wedge
\eta_{i}] +i_{\xi^{*(n+1)}}(dp_{\mu}\wedge dy^{\mu}\wedge
\eta)=\\
=[ \xi^{\mu}dp_{\mu}\wedge \eta+p_{\mu}d\xi^\mu \wedge \eta -\xi^{i}dp_{\mu}\wedge dy^\mu\wedge \eta_{i}-p_{\mu}d\xi^{i}\wedge dy^{\mu}\wedge \eta_{i}+p_{\mu}\xi^i \lambda_{G,i}dy^\mu \wedge \eta]+\\
+[\xi^{p_\mu}dy^{\mu}\wedge \eta -\xi^{\mu}dp_{\mu}\wedge \eta +\xi^{i}dp_{\mu}\wedge dy^{\mu}\wedge \eta_{i}]=
 [ \xi^{\mu}dp_{\mu}\wedge \eta+p_{\mu}\xi^{\mu}_{,y^\nu}dy^\nu \wedge \eta -\xi^{i}dp_{\mu}\wedge dy^\mu\wedge \eta_{i}+\\+p_{\mu}\xi^{i}_{,i}\wedge dy^{\mu}\wedge \eta+p_{\mu}\xi^i \lambda_{G,i}dy^\mu\wedge \eta]
+[\xi^{p_\mu}dy^{\mu}\wedge \eta -\xi^{\mu}dp_{\mu}\wedge \eta +\xi^{i}dp_{\mu}\wedge dy^{\mu}\wedge \eta_{i}]=\\
=[\xi^{p_{\mu}}+p_{\nu}\xi^{\nu}_{,y^\mu}+p_{\mu}(\xi^{i}_{,i}+\xi^{i}\lambda_{G,i})]dy^\mu\wedge \eta.
\end{multline}
Here we have used the relation $d\eta_{i}=\lambda_{G,i} \eta$.\par Equating the obtained expression
 to zero we get the expression for $\xi^{p^\mu}$ as in the Proposition.
\end{proof}

 Combining the last result with the prolongation $\xi^{*0}$
from the Definition-Proposition 10 and with the prolongation $\xi^{*(n+1)}$ from the previous Proposition
and using factorization by $\Lambda^{(n)+(n+1)}_{1}Y$ we get the following
\begin{corollary} For any projectable vector field  $\xi \in X(\pi)$ there exists unique projectable
vector field $\xi^{1*}$ in the space $\Lambda^{n+(n+1)}_{2}Y$ -
prolongation of $\xi$, \emph{preserving the $n+(n+1)$ form} $p\eta+ p^{\mu}_{\mu}dy^{\mu}\wedge
\eta_{i}+p_{\nu}dy^{\nu}\wedge \eta$.\par
Vector field $\xi^{1*}$ projects to $\Lambda^{(n+1)+(n+2)}_{2/1}Y$ to the vector field $\tilde{\xi}^{1}$ preserving the form $p^{\mu}_{\mu}dy^{\mu}\wedge
\eta_{i}+p_{\nu}dy^{\nu}\wedge \eta \ mod\ \eta.$
\end{corollary}

Below we will also need to define prolongations of projectable vector fields $\xi \in \mathcal{X}_{p}(\pi)$ to the space of the $n+(n+1)$-forms $\Lambda^{n+(n+1)}_{2/1}(J^{1}(\pi))$ on the 1-jet bundle $J^{1}(\pi)\rightarrow Y.$  Next statement, whose proof is presented in the Appendix 1, describes these prolongations. Local coordinates in this bundle corresponding to a fibred chart  $(W,x^i,y^\mu)$ is defined at the presentation of elements of the fibers:
\beq
p\eta +p^{\mu}_{i}dy^\mu \wedge \eta_i +q_\mu \omega^\mu \wedge \eta +q^{i}_{\mu}\omega^{\mu}_{i}\wedge \eta.
\eeq

\begin{proposition} For any projectible vector field $\xi=\xi^i \partial_{i}+\xi^\mu \partial_{\mu}+\xi^{\mu}_{i}\partial_{z^{\mu}_{i}} \in \mathcal{X}(\pi^{1}_{0})$ on the space $J^{1}_{p}(\pi)$, there exist unique vector field $\xi^{*}\in \mathcal{X}(\Lambda^{n+1}_{2}(J^{1}(\pi)))$ such that
\begin{enumerate}
\item Canonical $n+1$-form $Q^{n+1}=(q_\mu \omega^\mu +q^{i}_{\mu}\omega^{\mu}_{i})\wedge \eta$ (see ()) is invariant with respect to the flow of vector field $\xi^*$: $\mathcal{L}_{\xi^*}Q^{n+1}=0.$
\item This vector field is given by the following expression
\beq
\xi^{*} =\xi+(-q_\mu \xi^{\mu}_{y^\nu}-q^{i}_{\mu}\xi^{\mu}_{i,y^\nu}-q_\nu div_{G}(\bar \xi ))\partial_{q_\nu}+(-q_\mu \xi^{\mu}_{,z^{\nu}_{j}} -q^{i}_{\mu}\xi^{\mu}_{i,z^{\nu}_{j}} -q^{i}_{\nu} div_{G}(\bar \xi ))\partial_{q^{j}_{\nu}}.
\eeq
\end{enumerate}
\end{proposition}

\subsection{Partial higher order jet bundles $J^{k}_{p}(\pi)$.}
Here we introduce the higher order partial jet bundles defined by an AP-structure $T(X)=K\oplus K'$ and, more generally, by an $S$-structure. \textbf{In the rest of the paper we will understand by} $J^{k}_{p}(\pi),k=1,\ldots, \infty$ \textbf{one of the bundles} $J^{k}(\pi),J^{k}_{K}(\pi),J^{k}_{S}(\pi), J_{0}(\pi)$.  One can introduce more general notion of higher order (and infinite) partial jet bundle of a bundle $\pi$, but this will be done elsewhere.\par

Let $Z^{n}_{+}$ be the set of multiindices $\{I =(i_{1},\ldots ,i_{n})\vert i_{l}\in N\}$. Let $T(X)=K\oplus K'$ be an integrable AP-structure on $X$, and let $U,(x^j ,j =1,\ldots n_{K};x^k , k =n_{K}+1,\ldots n)$ be a (local) integrating chart in $X$. Distribution $K$ (respectively $K'$) in this chart is generated by vector fields $\partial_{x^j}$ (respectively, by vector fields $\partial_{x^k}$).  Let $N_{K}= \{(1,0,\ldots ,0),(0,1,\ldots ,0)\ldots (0,\ldots,0,1_{n_{K}},0,\ldots )\subset Z^{n}_{+} \}$ is the set of multiindices corresponding to the derivatives in $K$.  Finally let $\hat{N}_{K}=N_{K}+Z^{n}_{+}$ be the set of all multinidices larger or equal to the multiindices from $N_{K}$ with respect to the natural order in $Z^{n}_{+}$.\par
\begin{definition} Let $\pi_{XY}:Y\rightarrow X$ be a fiber bundle and let $K\subset T(X)$ be a subbundle.
\begin{enumerate}
\item Let $x\in U\subset X,$ $s_{1},s_{2}\in \Gamma (U,\pi)\vert s_{1}(x)=s_{2}(x)$. Sections $s_{1},s_{2}$ are called $K$-\textbf{equivalent of order k at a point} $x\in X:\ s_{1}\sim^{k}_{K_{x}} s_{2} $ if
$  \partial^{I}s_{1}(x)=\partial^{I}s_{2}(x),\ \forall I \in \hat{N}_{K}.$
\item  $J^{k}_{K\ x}(\pi)$- space of classes of $\sim^{k}_{K}$  at a point $x\in X$.
\item $J^{k}_{K}(\pi)=\cup_{x\in X}J^{k}_{K\ x}(\pi)$ - the space of partial k-jets of sections of $\pi$.
\item Infinite jet bundle $J^{\infty}_{K}(\pi)$ is the inverse limit of the bundles $J^{k}_{K}(\pi)$ under the natural projections $\pi_{k(k-1)}:J^{k}_{K}(\pi)\rightarrow J^{k-1}_{K}(\pi).$
\end{enumerate}
\end{definition}
Speaking simply, bundle $J^{k}_{K}(\pi)$ carry information about the derivatives $\partial_{x^j}s$ of section $s\in \Gamma(\pi)$ and all derivatives $\partial^{I}\partial_{x^j}s$ up to the order $k$. \par
In the next proposition we collect some basic properties of bundles $J^{1}_{K}(\pi)$ that will be used below.\par
\begin{proposition}
\begin{enumerate}
\item  Bundles $\pi_{k (k-1)}:J^{k}_{K}(\pi)\rightarrow J^{k-1}_{K}(\pi)$ form the inverse sequence of the affine bundles. \par
\item  There is a canonical surjection of affine bundles $w_{K}: J^{k}(\pi)\rightarrow  J^{k}_{K}(\pi)$
such that the diagram
\[
\begin{CD}
J^{k}(\pi)@> w_{K}>>  J^{k}_{K}(\pi)\\
@V\pi_{k (k-1)}VV   @V\pi_{k (k-1)}VV\\
J^{k-1}(\pi)@> w_{K}>>  J^{k-1}_{K}(\pi)\\
\end{CD}
\] is commutative.

\item Let $T(X)=K\oplus K'$ be an integrable AP-structure, $(x^j,x^k )$ - local chart integrating the $AP-structure$. The 1-forms
\[\begin{cases} \omega^{\mu}=dy^\mu-\sum_{j \in K}z^{\mu}_{j}dx^j;\ i\in \overline{1,m}, \\   \omega^{\mu}_{J }=dz^{\mu}_{J}-\sum_{l}z^{\mu}_{Jl}dx^{l} ,\ J \in \hat{N}_{K},\ 0< \vert J \vert < k,\ \mu\in \overline{1,m}
\end{cases}\]
generate the \emph{partial Cartan distribution} $C^{k}_{K}$ on the bundle $J^{k}_{K}(\pi )\rightarrow X$.
\item The "total derivative" operator $d_i=\partial_{x^i}+\sum_{(i,\mu)\in P}z^{\mu}_{i}\partial_{y^\mu}+\sum_{(I,\mu)\in P}z^{\mu}_{I+i}\partial_{z^{\mu}_{I}}$ is defined on the functions $C^{\infty}(J^{k}_{K}(\pi))$ and has the same properties as the usual total derivatives (commutativity, etc).  Here and below we use the notation $(\mu,I)\in P$ for the set of variables $z^{\mu}_{I},\vert I\vert > 0$ that are present in the partial k-jet bundle $J^{k}_{p}(\pi).$ The set $P$ if indices is invariant under the admissible automorphisms of $J^{k}_{p}(\pi)$
\item A section $q$ of the bundle $J^{k}_{K}(\pi )$ is the (partial) k-jet of a section $s:X\rightarrow Y$ if and only if $q^{*}(\omega^{\nu}_{J} )\vert_{K}=0$ for all 1-forms listed in p.3.
\item Any $\pi$-projectable vector field $\xi \in \mathcal{X}(Y)$ preserving the AP-structure $T(X)=K\oplus K'$ can be uniquely prolonged (by the flow prolongation) to the projectable vector field $\hat{\xi}^k \in X(J^{k}_{K}(\pi ))$ preserving the partial Cartan distribution $C^{k}_{K}$. For a vector field $\xi =\xi^i\partial_{x^i}+\xi^\mu \partial_{y^\mu}$ the k-th order prolongation has the form (see \cite{O}, Thm. 2.36)
    \beq
    pr^{(k)}\xi=\xi +\sum_{(\mu,I)\in P}\xi^{I}_{\mu}\partial_{z^{\mu}_{I}},\ \phi^{I}_{\mu}=d_I (\xi^\mu -\sum_{i}\xi^i z^{\mu}_{i}) +\sum_{i}\xi^i z^{\mu}_{I+i}.
    \eeq
    \item Contact decomposition (2.3) of the forms is valid for the tower of partial jet-bundles $J^{k}_{K}(\pi )$.
\end{enumerate}
\end{proposition}
Proof of almost all statements in this Proposition is straightforward, using the standard structural properties of the higher order jet bundles, see \cite{KMS,O}. In the last statement the prolongation is the standard flow prolongation of an admissible vector field, see Proposition 3 above. and Lemma before it.\par
For a space-time splitting $S$ of the space of fields $y^i$, see Sec.3.2, one can similarly define the bundles $J^{k}_{S}(\pi),J^{\infty}_{S}(\pi)$, their partial Cartan structure and establish the prolongation properties of $\pi$-projectable vector fields $\xi \in \mathcal{X}(Y)$ preserving the $S$-structure of the bundle $\pi$ and the existence of the contact decomposition of exterior forms.  We will use these results without further references.

\section{Constitutive relations (CR) and their Poincare-Cartan forms}
Here we introduce the constitutive and covering constitutive relations of order $k$ as generalized Legendre transformation from the
(partial) k-jet bundle $J^{k}_{p}(\pi )$ to the extended multisymplectic bundles of $n+(n+1)$-forms on the manifold $Y$,
 see next commutative diagram

 \beq  \bfig
\Vtriangle[J^{k}_{p}(\pi )` \Lambda_{2/1}^{n+(n+1)}Y`Y^{n+m};\mathcal{C}`\pi_{k0}`\pi^{n+(n+1)}]
\morphism(500,0)<0,-250>[Y^{n+m}`X^{n};\pi]
\morphism(1000,1000)<0,-500 >[\Lambda_{2}^{n+(n+1)}Y`\Lambda_{2/1}^{n+(n+1)}Y ;q]
\morphism(0, 500)<1000,500>[J^{k}_{p}(\pi )`\Lambda_{2}^{n+(n+1)}Y; \widehat{\mathcal{C}} ]
\efig
\eeq

\begin{definition}
\begin{enumerate}
\item  A \textbf{covering constitutive relation (CCR) $\hat C$ of order $k$} is morphism of bundles
$\hat{C}:J^{k}_{p}(\pi )\rightarrow \Lambda_{2}^{n+(n+1)}Y$ over $Y$:
\[\widehat{\mathcal{C}}(x^i ,y^\mu, z^{\mu}_{i},\ldots,z^{\mu}_{i_{1}\ldots i_{k}} ) =(x^i ,y^\mu;p;
F^{i}_{\mu};\Pi_{\mu}),\ p,F^{i}_{\mu},\Pi_{\mu}\in C^{\infty}(J^{k}_{p}(\pi )).
\]
 \textbf{The Poincare-Cartan form of the CCR} $\hat{C}$ is the form
\beq
\Theta_{\hat C}=\mathcal{\hat{C}}^{*}(\Theta^{n}_{2}+\Theta^{n+1}_{2})=p\eta+ F^{i}_{\mu}dy^i \wedge
\eta_{i}+\Pi_{\mu}dy^\mu \wedge \eta.
\eeq
\item  A \textbf{constitutive relation (CR) $C$ of order $k$} is a morphism of bundles $C:J^{k}_{p}(\pi )\rightarrow \Lambda_{2/1}^{n+(n+1)}Y$ over $Y$:
\[
C(x^i ,y^\mu, z^{\mu}_{i},\ldots,z^{\mu}_{i_{1}\ldots i_{k}} ) =(x^i ,y^\mu; F^{i}_{\mu};\Pi_{\mu}),\ F^{i}_{\mu},\Pi_{\mu}\in C^{\infty}(J^{k}_{p}(\pi )).
\]
\textbf{The Poincare-Cartan form of the CR} $\mathcal{C}$ is the form defined $ mod\  \eta$
\[
\Theta_{C}=\mathcal{C}^{*}(\Theta^{n}_{2}+\Theta^{n+1}_{2}\ mod\  \eta)= F^{i}_{\mu}dy^\mu \wedge
\eta_{i}+\Pi_{\mu}dy^\mu \wedge \eta\  mod\  \eta.
\]

\end{enumerate}
\end{definition}
When the order of a CR or CCR is one ($k=1$) we will omit the words "of order 1".  \par
Any covering constitutive relation defines the corresponding constitutive relation by the projection - $C=q\circ \hat{C}$.  On the other hand, there are many ways to lift a constitutive relation to the CCR.  One of them (for $k=1$) is to use the connection on the bundle $\pi$, see \cite{LM}.  In between all possible CCR corresponding to a given CR there is one privileged (as we will see in the next section)

\begin{definition}   The \textbf{lifted CCR} $\tilde C$ \textbf{of a constitutive relation} $C$ (of order $k$) is defined by
\beq
\Theta_{\tilde C}=F^{i}_{\mu}\omega^{\mu}\wedge \eta_{i}+\Pi_{\mu}\omega^{\mu}\wedge \eta =-(\sum_{(\mu,j )\in P}z^{\nu}_{j}F^{j}_{\nu})\eta +F^{i}_{\mu}dy^\mu \wedge
\eta_{i} + \Pi_{\mu}dy^\mu \wedge \eta.
\eeq
\end{definition}
As the next Lemma shows, the Poincare-Cartan form of the lifted CCR can be obtained with the help of the vertical endomorphism $S_{\eta}$.  Proof of this Lemma is straightforward.
\begin{lemma} Let $k=1$.  For a 1-form $\chi =F^{i}_{\mu}dz^{\mu}_{i}+\Pi_{\mu}dy^\mu\in \Lambda^{1}(J^{1}(\pi )),$ one has \[ \Theta_{{\tilde C}}=S_{\eta}^{*}(F^{i}_{\mu}dz^{\mu}_{i}+\Pi_{\mu}dy^\mu ).\]
\end{lemma}
Remind, that the Poincare-Cartan form $\Theta_{L}$ of the the Lagrangian form $L\eta$ is an example of the \emph{Lepage form} corresponding to the Lagrangian $L$ (ref). Having defined the Poincare-Cartan form of a balance system with the covering constitutive relations $\hat{C}$ it is interesting to see when such a form may be a Lepage form and if yes, to which Lagrangian it corresponds. Next result provides the answer to this question.

\begin{lemma} The term $\Theta^{n}_{\hat C}$ of the Poincare-Cartan form (9.4) of a covering constitutive relation $\hat C$ is the \emph{Lepage form} in $J^{k}(\pi)$ if and only if $\hat C$ is the CCR of order one, i.e. that $p,F^{i}_{\mu}\in \pi^{k*}_{1}C^{\infty}(J^{1}(\pi))$. If this condition is fulfilled, the associated Lagrangian of the Lepage form $\Theta^{n}_{\hat C}$ is equal to  $L =p+\sum_{(\mu ,i)\in P}z^{\mu}_{i}F^{i}_{\mu}.$
\end{lemma}
\begin{proof} To prove the first statement we write the form $\Theta^{n}_{\hat C}$ as follows
\[
\Theta^{n}_{\hat C}=p\eta +F^{i}_{\mu}dy^\mu \wedge \eta_{i}=(p+z^{\mu}_{i}F^{i}_{\mu})\eta +F^{i}_{\mu}\omega^\mu \wedge \eta_{i},
\]
and denote $\tilde p =p+z^{\mu}_{i}F^{i}_{\mu}$.  Now we calculate
\begin{multline*}
\pi^{(k+1)*}_{k}d\Theta^{n}_{\hat C}= \pi^{(k+1)*}_{k}(d\tilde{p}\wedge \eta +dF^{i}_{\mu}\wedge \omega^\mu \wedge \eta_{i}+F^{i}_{\mu}d\omega^{\mu}\wedge \eta_{i} -F^{i}_{\mu}\lambda_{G,i} \omega^\mu \wedge \eta)=\\=
(d_{h}{\tilde p}+d_{v}\tilde p )\wedge \eta+(d_{h}F^{i}_{\mu}+d_{v}F^{i}_{\mu})\wedge \omega^{\mu}\wedge \eta_{i}-F^{i}_{\mu}dz^{\mu}_{j}\wedge dx^j \wedge \eta_{i}-F^{i}_{\mu}\lambda_{G,i} \omega^\mu \wedge \eta) =\\
=(\sum_{(\nu,I),\vert I \vert \geqq 0}{\tilde p}_{,z^{\nu}_{I}}\omega^{\nu}_{I} )\wedge \eta+d_{j}F^{i}_{\mu}dx^j \wedge \omega^{\mu}\wedge \eta_{i}+(\sum_{(\nu,I),\vert I \vert \geqq 0}F^{i}_{\mu, z^{\nu}_{I}}\omega^{\nu}_{I})\wedge \omega^{\mu}\wedge \eta_{i}-F^{i}_{\mu}dz^{\mu}_{i}\wedge \eta-F^{i}_{\mu}\lambda_{G,i} \omega^\mu \wedge \eta=\\=-(d_{i}F^{i}_{\mu}+F^{i}_{\mu}\lambda_{G,i})\omega^{\mu}\wedge \eta+ {\tilde p}_{,y^\nu}\omega^\nu \wedge \eta + ({\tilde p}_{,z^{\mu}_{i}}-F^{i}_{\mu})\omega^{\mu}_{i}\wedge \eta+
\sum_{(\nu,I),\vert I \vert > 1}{\tilde p}_{,z^{\nu}_{I}}\omega^{\nu}_{I} \wedge \eta +2Con.
\end{multline*}
In the last expression first four terms present 1-contact part, the last one - 2-contact part $(\sum_{(\nu,I),\vert I \vert \geqq 0}F^{i}_{\mu, z^{\nu}_{I}}\omega^{\nu}_{I})\wedge \omega^{\mu}\wedge \eta_{i}$.  1-contact part is $\pi^{k+1}_{0}$-horizontal if and only if conditions
\[
\begin{cases}
{\tilde p}_{,z^{\nu}_{I}}=0,\ \vert I \vert > 1,\\
{\tilde p}_{,z^{\mu}_{i}}-F^{i}_{\mu}=0,\ \forall (\mu,i).
\end{cases}
\]
First condition requires that ${\tilde p}\in \pi^{k*}_{1}C^{\infty}(J^{1}(\pi))$.  If this fulfilled, then the second condition requires that $F^{i}_{\mu}\in  \pi^{k*}_{1}C^{\infty}(J^{1}(\pi))$ and therefore the CCR $\hat C$ is of the order one. What is left is that the second condition tells that the density/flux part $F^{i}_{\mu}$ components of constitutive relations come from the "Lagrangian" $\tilde p$.  Below such constitutive relations will be called "semi-Lagrangian".
\end{proof}
\subsection{Examples.}
Present now several examples of different types of constitutive relations or covering constitutive relations and corresponding Poincare-Cartan forms.  In all these examples $k=1$.
\begin{example} A \textbf{Lagrange constitutive relation of order 1} is defined by a function
$L\in C^{\infty}(J^{1}(\pi))$:
\[ C_{L}(x^i ,y^\mu, z^{\mu}_{i})=(p^{i}_{\mu}=F^{i}_{\mu}=\frac{\partial L}{\partial z^{\mu}_{i}};\Pi_{\mu}=\frac{\partial L}{\partial y^{\mu}}). \]
Balance system for such a constitutive relation (see next section) coincide with the conventional system of Euler-Lagrange Equations (2.10) of the first order Lagrangian Field Theory.
\end{example}
\begin{remark} For $k=1$ Euler-Lagrange Equations (2.10) have the form of a balance system with canonically defined densities and flux components.  For $k>1$ Euler-Lagrange equations can also be written as a balance system but in this case there are different ways to specify flux and even density components due to the presence of higher derivatives. This non-uniqueness is similar to the non-uniqueness of a choice of Lepage form for higher order Lagrangian Variational Theory, see \cite{FF}. That why in the Lagrangian case we take $k=1$ only.
\end{remark}
\begin{example} A \textbf{semi-Lagrangian CCR} $\hat{C}_{L,Q}$ is defined by a functions $L,Q_{\mu}, \mu=1,\ldots ,m\ \in
C^{\infty}(J^{1}(\pi))$:
\[
\hat{C}_{L,Q_{\mu}}(x^i ,y^\mu, z^{\mu}_{i})=(p=L-z^{\mu}_{i}L_{,z^{\mu}_{,i}},p^{i}_{\mu}=\frac{\partial L}{\partial
z^{\mu}_{i}};\Pi_{\mu}=Q_{\mu}(x^i ,y^\mu, z^{\mu}_{i})).
\]
\end{example}
\begin{example} \textbf{$L+D$-system}.  Let $L\ \text{Lagragian},D\ (\text{"dissipative potential"})\in C^{\infty}(J^{1}_{p}(\pi)).$ Let the time derivatives of the fields $y^\mu$ - $z^{\mu}_{0}$ are present in the partial 1-jet bundle $J^{1}_{p}(\pi).$ The \textbf{CR} $\mathcal{C}_{L,D}$ is defined by its Poincare-Cartan form
\[ \Theta_{L,D}=L_{z^{\mu}_{i}}dy^\mu \wedge \eta_{i}+(D_{z^{\mu}_{,0}}-L_{,y^\mu})dy^\mu \wedge \eta. \]
balance system for such constitutive relation have the form
\[\frac{\delta L}{\delta y^\mu}=\frac{\partial D}{\partial \dot{y}^\mu}\]
which is well known in the Continuum Thermodynamics, \cite{Ma1, Ma2}.
\end{example}
\begin{remark}
If we would like to define directly the analog of Lagrangian or semi-Lagrangian relation on a partial 1-jet bundle we would have a situation where an absence of a variable $z^{\mu}_{i}$ from the partial 1-jet bundle leads to the nullity of the corresponding flux component. This restricts an application of a conventional semi-Lagrangian CR defined on a partial jet bundles.  On the other hand, it is possible that the CR are \emph{partly variational} in the sense that part of the flux components are defined by a "partial Lagrangian" while other components are to be defined independently, see below, Prop.10.
\end{remark}
\begin{example} Consider a case where only spacial but not time derivatives of fields $y^\mu$ enters the constitutive relations: $J^{1}_{p}(\pi)=J^{1}_{T(B)}(\pi)$. Then it is possible that for a function $L\in C^{\infty}(J^{1}_{T(B)}(\pi))$, $F^{A}_{\mu}=\frac{\partial L}{\partial z^{\mu}_{A}},\ A=2,\ldots ,n;\ \Pi_{\mu}=\frac{\partial L}{\partial y^\mu}$ but densities $F^{0}_{\mu}$ are to be defined independently. Then,
\[
\Theta_{C}=F^{0}_{\mu}dy^\mu \wedge \eta_{0}+\frac{\partial L}{\partial z^{\mu}_{A}}dy^\mu\wedge \eta_{A}+\frac{\partial L}{\partial y^\mu}dy^\mu\wedge \eta.
\]
The system of balance equations corresponding to this CR is, of course, first order by time derivatives.\par
Five fields fluid system (Sec.2.7) is an example of such balance system.
\end{example}
\begin{example} \textbf{Vector-potential CR}. (RET case, dual variables) In this case $J^{1}_{p}(\pi)=\{ \cdot \}$ has a zero dimensional (point) fiber over $Y$.  Let $h=h^i (x,y)\eta_{i}$ be a semi-basic (n-1)-form on $Y$. Define the \textbf{CR}
\[ C_{h}(x^i ,y^\mu)=(p^{i}_{\mu}=\frac{\partial h^i}{\partial y^\mu};\Pi_{\mu}=\Pi_{\mu}(x,y)). \]
\end{example}
\subsection{Form $K_C$ and the Helmholtz coindition}
Let $C$ be a constitutive relations of first order with the lifted Poincare-Cartan form $\Theta_{\widetilde{C}}=F^{i}_{\mu}\omega^\mu \wedge \eta_{i}+\Pi_{\mu}\omega^{\mu}\wedge \eta$.  Associate with this constitutive relation the following contact (n+1)-form on $J^{2}(\pi)$.
\beq
K_{C}=(F^{i}_{\mu}\omega^{\mu}_{i}+\Pi_{\mu}\omega^\mu )\wedge \eta.
\eeq
Calculate differential of this form:
\beq
dK_{C}=[(\partial_{y^\nu}\Pi_{\mu})dy^{\nu}\wedge dy^{\mu}+(\partial_{y^\nu}F^{\mu}_{i}-\partial_{z^{\nu}_{j}}\Pi_\mu )dy^{\nu}\wedge dz^{\mu}_{i} +(\partial_{z^{\nu}_{j}}F^{\mu}_{i})dz^{\nu}_{j}\wedge dz^{\mu}_{i} ]\wedge \eta.
\eeq
Now we use the following form of Poincare Lemma
\begin{lemma} Let $f_i  , i=1,\ldots ,k; g_j,\ j=1,\ldots ,s$ be functions of all variables $z^i,  i=1,\ldots ,k;\ y^j,\ j=1,\ldots ,s$ such that the form $K=f_{i}dz^i+g_{j}dy^j $ is closed: $dK=0$.  Then, locally, $K=dL,$ i.e. $f_{i}=\partial_{z^i}L,\ g_j =\partial_{y^j}L$ for some smooth function $L=L(z^i ,y^j )$.
\end{lemma}
Applying this Lemma to the second form of equality (4.5) we get the following version of (local) Helmholtz condition (comp. \cite{KS}) for a balance system of order one to be Euler-Lagrange system for some Lagrangian $L\in C^{\infty}(J^{1}(\pi))$.
\begin{proposition} For a constitutive relations $C$ of order 1 the following properties are equivalent:
\begin{enumerate}
\item Form $K_{C}$ is closed: $dK_{C}=0$,
\item $C$ is locally Lagrangian constitutive relation: $C=C_{L}$ for a (locally defined) function $L\in C^{\infty}(J^{1}(\pi))$.
\end{enumerate}
\end{proposition}
\begin{remark} Construction of the form $K_C$ is related to the mapping $\omega^{\mu}\wedge \eta_{i}\rightarrow \omega^{\mu}_{i}\wedge \eta$ of $J^{1}(\pi)\rightarrow J^{2}(\pi)$.  It would be interesting to construct such a mapping from $J^{k}(\pi)$ to $J^{k+1}(\pi)$ for $k>1$.
\end{remark}
\subsection{Variational sequence and the balance systems}
Condition of the Proposition 8 for a balance system to be Lagrangian points to the possibility to interpret and study the balance systems using the variational bicomplex, see \cite{An,O}, or the variational sequence, \cite{V}. In this section we realize this possibility.  In presenting the variational bicomplex we will follow \cite{V}.\par

We will be using the augmented variational bicomplex in the form

\beq
\begin{CD}
@. 0 @. 0 @.@.@.\\
 @.  @VVV @VVV @. @. @. \\
@. R @. R @. 0 @. 0 @. 0\\
@. @VVV @VVV @VVV @VVV @VVV \\
0 @>>> \Omega^{0}(M) @>\pi^{*}_{\infty}>> E^{0,0}_{0} @>d_v >> E^{1,0}_{0} @>.......>> E^{p,0}_{0} @>d_v >>E^{p+1,0}_{0} @>...>> \\
 @. @VdVV @Vd_hVV @V-d_h VV @V (-1)^p d_h VV  @V (-1)^{p+1} d_h VV  \\
0 @>>> \Omega^{1}(M) @>\pi^{*}_{\infty}>> E^{0,1}_{0} @>d_v >> E^{1,1}_{0} @>.......>> E^{p,1}_{0} @>d_v >>E^{p+1,1}_{0} @>...>> \\
@.        @V...VV     @V...VV @V...VV  @V...VV  @V...VV \\
0 @>>> \Omega^{n-1}(M) @>\pi^{*}_{\infty}>> E^{0,n-1}_{0} @>d_v >> E^{1,n-1}_{0} @>.......>> E^{p,n-1}_{0} @>d_v >>E^{p+1,n-1}_{0} @>...>>\\
@. @VdVV @Vd_hVV @V-d_h VV @V (-1)^p d_h VV  @V (-1)^{p+1} d_h VV \\
0 @>>> \Omega^{n}(M) @>\pi^{*}_{\infty}>> E^{0,n}_{0} @>d_v >> E^{1,n}_{0} @>.......>> E^{p,n}_{0} @>d_v >>E^{p+1,n}_{0} @>...>>\\
@. @VVV @V\pi VV @VIVV @VIVV @VIVV \\
@. 0 @>>> E^{0,n}_{1} @>e_1 >> E^{1,n}_{1}  @>...... >> E^{p,n}_{1} @>e_1 >> E^{p+1,n}_{1} @>....>> \\
@. @. @VVV @VVV @VVV @VVV \\
@. @. 0 @. 0 @. 0 @. 0
 \end{CD}
\eeq
In this diagram, $E^{0,q}_{0}=\Omega^{0,q}(J^{\infty}(\pi))$ - the bundle of smooth horizontal $q$-forms on $J^{\infty}(\pi)$, $E^{p,q}_{0}=C^{p}\bigwedge \Omega^{0,q}(J^{\infty}(\pi))$ is the bundle of $p$-vertical and $q$-horizontal $p+q$-forms. Terms of the lowest line represent the fist derived complex
\[
E^{p,n}_{1}=E^{p,n}_{0}/d_{h}(E^{p,n-1}_{0})=C^{p}\bigwedge \Omega^{0,n}/d_{h}(C^{p}\bigwedge \Omega^{0,n-1}).
\]
The mappings $e_1 :E^{p,n}_{1}\rightarrow E^{p+1,n}_{1},\ e_{1}([\alpha])=[d_v \alpha ]$ participate in the last horizontal line of the bicomplex as well as in the second half of the induced  variational sequence (where Euler-Lagrange mapping $\mathcal{E} =e_1 \circ \pi$, $\pi$ being the quotient projection defining $E^{0,n}_{1}$)

 \beq
 0 \rightarrow \mathbf{R} \rightarrow E^{0,0}_{0} \xrightarrow{d_h}  E^{1,0}_{0}  \xrightarrow{d_h} \ ... \  \
 E^{0,n-1}_{0} \xrightarrow{d_h}  E^{0,n}_{0} \xrightarrow{\mathcal{E}} E^{1,n}_{1} \xrightarrow{e_1} E^{2,n}_{1} ............. E^{p,n}_{1} \xrightarrow{e_1} E^{p+1,n}_{1}\cdots
 \eeq
Operator $I$ is the \emph{interior Euler}operator $I:\Omega^{s,n}(J^{\infty}(\pi)) \rightarrow \Omega^{s,n}(J^{\infty}(\pi))$, defined by (\cite{An})
\beq
I(\omega) =\frac{1}{s}\omega^i \wedge \left[ i_{\partial_{i}}\omega -d_{\mu}(i_{\partial_{z^{i}_{\mu}}}\omega +d_{\mu_1 \mu_2}(i_{\partial_{z^{i}_{\mu_1 \mu_2}}}\omega ) -\ldots \right].
\eeq
Interior euler operator is closely related to the Euler-Lagrange operator $\mathcal{E}$, namely (see \cite{An}, Ch. 2), for any Lagrangian n-form $\lambda =L\eta$
\beq
E(\lambda)=I\circ d_v (\lambda).
\eeq
That is why the next proposition is hardly surprising
\begin{proposition}
Let $\mathcal{C}$ be a CR and $K_\mathcal{C}=\Pi_{i}\omega^{i}\wedge \eta +F^{\mu}_{i}\omega^{i}_{\mu}\wedge \eta$ - corresponding K-form introduced in (4.4).  Then, the balance system $\bigstar$ is equivalent to the equation
\beq
j^{*}s i_{\xi^1}I(K_{\mathcal{C}})=0,
\eeq
for all $\xi \in \mathcal{V}(\pi)$.\par
\end{proposition}
\begin{proof}
\beq
I(K_{\mathcal{C}})= \omega^i \wedge [\Pi_{i}\eta -d_{\mu} (F^{\mu}_{i}\eta)] = \omega^{i}\wedge [\Pi_{i}  -d_{\mu}F^{\mu}_{i} -F^{\mu}_{i}\lambda_{G,\mu}]\eta .
\eeq
Here we have used the form $\eta=\sqrt{\vert G\vert } dx^1 \wedge \ldots \wedge dx^n$.\par
Applying $i_{\xi^1}$ for any variational vector field $\xi \in \Gamma (\mathcal{V}(\pi))$ and taking pullback by the mapping $J^{*}s$ we get equation () in the form
\[
\omega^{i}(\xi)\circ j^{1}s [  d_{\mu}F^{\mu}_{i} +F^{\mu}_{i}\lambda_{G,\mu}-\Pi_{i}]\circ j^{\infty}s \eta =0.
\]
Fulfillment of this equality for all (or many enough) variational vector fields $\xi$. is equivalent to the statement that $s$ is the solution of the balance system $\bigstar$.
\end{proof}

\section{Variational form of balance systems}
In this section we present the invariant variational form of balance system and the separation of this invariant form into the $m$ separate balance laws by independent variations.  In difference to the Lagrangian case, in general, one has to put a condition on the variations that can be used for separating equations.  In cases of semi-Lagrangian, order 1 or RET balance systems that condition is fulfilled for all variations as in the conventional Lagrangian theory.  In general this situation can be remedied by modifying the source term $\Pi_{\mu}dy^\mu \wedge \eta$ - addition to it some contact form (see below, Sec.5.2).  We prove the main result for the case of a constitutive relation of arbitrary finite order, so, for simplicity we will consider that a CR $\mathcal{C}$ is defined on the infinite partial jet bundle $J^{\infty}_{p}(\pi)$ but is $\pi^{\infty}_{k}$-projectable for some $k<\infty$.

\subsection{Invariant form of balance systems}
We start with an arbitrary covering constitutive relation $\widehat C$ and change the sign of source term, i.e. we consider
\[ \Theta_{\widehat{ \mathcal{C_{-}}}}=
p\eta +F^{i}_{\mu}dy^\mu \wedge \eta_{i}-\Pi_{\mu}dy^\mu\wedge \eta .\]
For a section $s\in \Gamma_{V}(\pi ),\ V\subset X$ we request the fulfilment of the equation (Invariant Balance System, shortly IBS)
 \begin{equation} j^{k}(s)^{*}(i_{\xi}{\tilde d}\Theta_{{\hat C}_{-}})=0 ,\ \xi \in \mathcal{X}(J^{k}_{p}(\pi ))\ \ \ \ \ \  (IBS) \end{equation}
for a large enough family of variations $\xi$ - sufficient for separation of individual balance laws (see Def.7 below).\par
 Thus, we take the $n+(n+1)$-form $\Theta_{{\hat C}_{-}}$ of the form (4.2) and apply
first the Con-differential ${\tilde d}$ and then $i_{\xi}$ for a vector field $\xi
=\xi^{j}\partial_{j}+ \xi^{\nu}\partial_{y^{\nu}}+\sum_{\vert I \vert >1}\xi^{\mu}_{I}\partial _{z^{\mu}_{I}}$.\par
  Adopting the summation by
repeated indices we recall that only $z^{\mu}_{I}$ or derivatives by these variables with $(\mu,I )\in P$ are present in the formulas. We introduce the notation  $d\eta_{i}=\lambda_{G,i}\eta$.  It will be convenient to include variables $y^\mu$ into the family of variables $z^{\mu}_{I}$ for $\vert I \vert =0$ taking $z^{i}=y^i$.\par
We will also use the contact splitting (2.3) of the lift of differential of a function $p$ to $J^{\infty}_{p}(\pi)$ and similar contact decomposition for the flux components $F^{j}_{\nu}$.\par
We have,
\beq
{\tilde d}\Theta_{{\hat C}_{-}}=d(p\eta +F^{i}_{\mu}dy^\mu \wedge \eta_{i})+\Pi_{\mu}dy^\mu\wedge \eta =
dp\wedge \eta +dF^{i}_{\mu}\wedge dy^\mu \wedge \eta_{i}-F^{\mu}_{\mu}dy^\mu \wedge \lambda_{G,i}\eta +\Pi_{\mu}dy^\mu\wedge \eta =
\eeq
Now we continue the calculation replacing in the last two terms  in (5.2) the $dy^\mu$ by $\omega^\mu$ and using $dy^\nu \wedge \eta_{j}=\omega^\nu \wedge \eta_{j}+z^{\nu}_{k}dx^k \wedge \eta_{j}=\omega^\nu \wedge \eta_{j}+z^{\nu}_{j}\eta$
\begin{multline}
=[(d_{i}p)dx^i -\sum_{(\mu,I )\in \hat{P}}p_{,z^{\mu}_{I }}\omega^{\mu}_{I}]\wedge \eta +\\+[(d_{i}F^{j}_{\nu})dx^i -\sum_{(\mu,I)\in \hat{P}}({F^{j}_{\nu}})_{,z^{\mu}_{I }}\omega^{\mu}_{I}]\wedge dy^\nu \wedge \eta_{j}-\lambda_{G,i}F^{i}_{\mu}\omega^\mu \wedge \eta +\Pi_{\mu}\omega^\mu\wedge \eta=\\
=(d_{i}F^{j}_{\nu})dx^i \wedge dy^\mu \wedge \eta_{j}-\lambda_{G,i}F^{i}_{\mu}\omega^\mu \wedge \eta +\Pi_{\mu}\omega^\mu\wedge \eta
-[\sum_{(\mu,I )\in \hat{P}}p_{,z^{\mu}_{I}}\omega^{\mu}_{I }]\wedge \eta +\\
-[\sum_{(\mu,I )\in \hat{P}}({F^{j}_{\nu}})_{,z^{\mu}_{I }}\omega^{\mu}_{I }]\wedge (\omega^j \wedge \eta_{j}+z^{\nu}_{j}\eta)=\\
=(-d_{j}F^{j}_{\nu})\omega^\nu \wedge \eta-\lambda_{G,i}F^{i}_{\mu}\omega^\mu \wedge \eta +\Pi_{\mu}\omega^\mu\wedge \eta
-\sum_{(\mu,I )\in \hat{P}}[ p_{,z^{\mu}_{I}}+z^{\nu}_{j}{(F^{j}_{\nu})}_{,z^{\mu}_{I }} ]\omega^{\mu}_{I }\wedge \eta +2Con=\\
=[-d_{\mu}F^{\mu}_{i})-\lambda_{G,\mu}F^{\mu}_{i} +\Pi_{i}]\omega^i\wedge \eta +F^{\nu}_{j}\omega^{j}_{\nu}\wedge \eta +
[\sum_{(\mu,I )\in \hat{P}} (p+z^{\nu}_{j}F^{j}_{\nu})_{,z^{\mu}_{I}}\omega^{\mu}_{I }\wedge \eta +2Con.
\end{multline}
Here we have used the equality $(d_{i}F^{j}_{\nu})dx^i \wedge dy^\mu \wedge \eta_{j}=-(d_{j}F^{j}_{\nu}) dy^\mu \wedge \eta=-(d_{j}F^{j}_{\nu}) \omega^\mu \wedge \eta$.  Term $2Con$ in the last formula represents 2-contact form $[\sum_{(\mu,I )\in P}({F^{j}_{\nu}})_{,z^{\mu}_{I}}\omega^{\mu}_{I }]\wedge \omega^j \wedge \eta_{j}.$\par
Last formula proves the first statement of the next
\begin{theorem} Let $\hat C$ be a CCR defined in a domain of the partial k-jet bundle $J^{k}_{p}(\pi)$, $k\geqq 1$. Then,
\begin{enumerate}
\item We have the following contact decomposition
\begin{multline}
{\tilde d}\Theta_{{\hat C}_{-}}=[-d_{i}F^{i}_{\mu})-\lambda_{G,i}F^{i}_{\mu} +\Pi_{\mu}]\omega^\mu \wedge \eta +F^{j}_{\nu}\omega^{\nu}_{j}\wedge \eta
-\\ -\left[\sum_{(\mu,I)\in \hat{P}} (p+\sum_{(\nu,j)\in P}z^{\nu}_{j}F^{j}_{\nu})_{,z^{\mu}_{I}}\omega^{\mu}_{I}\right]\wedge \eta +2Con.
\end{multline}
Term in the brackets is the \textbf{vertical differential} $d_{v}(p+z^{\nu}_{j}F^{j}_{\nu})$ of the function $p+z^{\nu}_{j}F^{j}_{\nu}$, see \cite{GMS,KV}.  Internal sum is taken over all indices $(\nu,j)$ such that $z^{\nu}_{j}\in J^{1}_{p}(\pi)$ is in the 1-jet projection of $J^{k}_{p}(\pi)$, while the outside sum is taken over all $(\mu,I )$ with $z^{\mu}_{I}$ in $J^{k}_{p}(\pi)$.
\item  Let
$\xi \in \mathcal{X}(J^{k}_{p}(\pi))$ be any vector field. Then,
\beq i_{\xi}{\tilde d}\Theta_{{\hat C}_{-}}=- \omega^{1}_{\widehat{C}}(\xi)\eta -\omega^{k+1}_{\widehat{C}}(\xi^{k+1})\eta +Con,
\eeq
for an arbitrary prolongation $\xi^{k+1}$ to the vector field $\xi$ to $J^{k+1}_{p}(\pi)$, where
 \beq \begin{cases}
\omega^{1}_{\hat C}(\xi)=\omega^{\mu}(\xi)[d_{i}F^{i}_{\mu}+\lambda_{G,i}F^{i}_{\mu}-\Pi_{\mu}],\\
\omega^{k+1}_{\hat C}(\xi^{k+1}) =\sum_{(\nu,j)\in P}F^{j}_{\nu}\omega^{\nu}_{j}(\xi^{k+1} )
-\left[\sum_{(\mu,I )\in \hat{P}} (p+\sum_{(\mu,j)\in P}z^{\nu}_{j}F^{j}_{\nu})_{,z^{\mu}_{I }}\omega^{\mu}_{I }(\xi^{k+1})\right].
\end{cases}\eeq
Here $\xi^2$ is the projection of $\xi$ to $J^{2}_{p}(\pi)$ for $k\geqq $ and the prolongation to this subbundle for $k=1$.
\item For the lifted CCR $\tilde C$ of a constitutive relation $C$ and an arbitrary vector field $\xi \in \mathcal{X}(J^{k}_{p}(\pi))$,
\beq
 i_{\xi}{\tilde d}\Theta_{{\tilde C}_{-}}=- \omega^{1}_{\tilde {C}}(\xi)\eta -\sum_{(\nu,j)\in P}F^{j}_{\nu}\omega^{\nu}_{j}(\xi^{k+1} )\eta +Con
\eeq
\item Decompositions (5.4,5.7) are covariant with respect to a change of admissible coordinate
system $(x^i ,y^\mu ) \rightarrow (x'^{i}= h^i(x^j ), y'^{\mu}= h^\mu (x^j ,y^\nu )$ preserving the
"partial" structure (both in $K\oplus K'$ and $S$ cases).
\end{enumerate}
\end{theorem}
\begin{proof} Previous discussion proves the validity of the first decomposition. Second and third statements
follow from the first one.\par
Last statement follows from the tensorial behavior of both components of decompositions with regard of
such coordinate changes.\end{proof}
\begin{remark} Quantities
$Q_{\mu}=\omega^{\mu}(\xi)= \xi^{\mu}-z^{\mu}_{i}\xi^{i}$ form the \emph{characteristic of the vector field}
$\xi =\xi^{i} \partial_{i}+ \xi^{\mu}\partial_{y^\mu}$ in the sense of \cite{O}, Ch.2, or the \emph{generating section} in terms of \cite{KV}. \end{remark}

\begin{remark} Equality (5.7) contains the jet variables of the second order $z^{\mu}_{i I}$.  Yet,
\textbf{only} $\omega^{\mu}_{i}$ with $(i ,\mu)\in P$ \textbf{are present in the formula} (5.7)!  For
instance in the RET case \emph{all these terms are absent} from (5.7).\par

It is seen from the formula (5.7) that the lifted covering constitutive relation is privileged in the sense that the formula for ${\tilde d} \Theta_{\hat C_{-}}$ simplifies essentially for lifted CCR ${\tilde C}_{-}$.\par
In order that the equation resulting from taking the pullback by $j^{k}(s)$  in (5.4) \textbf{would not depend on the
variables not in} $J^{k}_{p}(\pi)$ we require that all the coefficients of these variables would be zero.
This leads to the condition one has to put to the allowed variations $\xi$ (see Def.7 below).
\end{remark}

Consider now two cases where no restrictions to the variations of Poincare-Cartan form appears.
\begin{proposition}  Let $\hat C$ be a CCR \textbf{of order k}. Then,
\beq \omega^{k+1}_{\hat C} =\sum_{(\nu,j)\in P}F^{j}_{\nu}\omega^{\nu}_{j}
-\left[\sum_{(\mu,I )\in \hat{P}} (p+\sum_{(\nu,j)\in P}z^{\nu}_{j}F^{j}_{\nu})_{,z^{\mu}_{I }}\omega^{\mu}_{I }\right]\equiv 0 \eeq
if and only if for some $L\in C^{\infty}(J^{1}_{p}(\pi))$,  $F^{i}_{\mu}=\partial_{z^{\mu}_{i}}L,\ (\mu,i)\in P.$ i.e. CR $\mathcal{C}$ is \textbf{(locally) semi-P-Lagrangian (lagrangian by variables $z^{\mu}_{i}\vert (i,\mu)\in P$ }) and \emph{does not depend on the variables }$z^{\mu}_{I},\vert I \vert >1$. In this case, $p=L-\sum_{(\mu,i)\in P}z^{\mu}_{i}\partial_{z^{\mu}_{i}}L +l(x,y)$ with an arbitrary $l(x,y)\in C^{\infty}(Y).$
\end{proposition}
\begin{proof}
Combining terms with $\vert I \vert =1$ in the sum with the first term in (5.8) and using linear independence of basic contact forms we split condition (5.8) into the following two groups of conditions
\[
\begin{cases}
1.\ F^{j}_{\nu}-\partial_{z^{\nu}_{j}}(p+\sum_{\mu,k )\in P}z^{\mu}_{k}F^{k}_{\mu})=0,\ \forall\ (j,\nu)\in P,\\
2.\  (p+ \sum_{(j,\nu)\in P}z^{\nu}_{j}F^{j}_{\nu})_{,z^{\mu}_{I }}=0,\ \forall\ (\mu,I)\in \tilde P,\ \vert I \vert >1.
\end{cases}.
\]
 Rewrite the equalities of the first group in the form
 \beq
\partial_{z^{\mu}_{j}}p=-z^{\mu}_{k}\partial_{z^{\mu}_{j}}F^{k}_{\mu}.
\eeq To see that $\Rightarrow $ holds we notice that provided left statement is true, the right sides of these
equalities satisfy to the mixed derivative test
\[
\partial_{z^{\nu}_{k}}(-z^{\lambda}_{i}\partial_{z^{\mu}_{j}}F^{i}_{\lambda})=
\partial_{z^{\mu}_{j}}(-z^{\lambda }_{i}\partial_{z^{\nu}_{k}}F^{i}_{\lambda}),
\]
or $\partial_{z^{\nu}_{k}}F^{l}_{\mu}=\partial_{z^{\mu}_{l}}F^{k}_{\nu}.$  It follows from this
equality valid for all couples of indices  $(\nu,k ),(\mu,l )\in P$ that there exists a function $L\in C^{\infty}(J^{k}_{p}(\pi))$ such that $F^{j}_{\nu}=\partial_{z^{\nu}_{j}}L,\forall (\nu,j) \in P$.  Substituting this to the first family of equalities we see that the function
$p+\sum_{\mu,k )\in P}z^{\mu}_{k}L_{,z^{\mu}_{k}}-L$ does not depend on $z^{\mu}_{i},(\mu,i)\in P.$ Therefore,
\[ p=L-\sum_{(\mu,k )\in P}z^{\mu}_{k}L_{,z^{\mu}_{k}}+l(x,y,z^{\nu}_{l},\ (\mu,l)\notin P).
\]
Moving sum in the right side to the left  and using second group of equations (5.9) we see that the function $L+l$ does not depend on the variables $z^{\nu}_{I},\ (\nu,I)\in \hat{P}\smallsetminus P.$
Using this function instead of $L$ we get the conclusion of Proposition.\par
To prove the opposite - reverse the arguments.
\end{proof}
Notice that the conclusion of this proposition put no restrictions on the components $F^{\mu}_{i}, (\mu,i)\notin P$.\par
In the case of a semi-Lagrangian CR there are no restriction to the variations in the IBS equation (5.1) provided one uses the proper lift of the constitutive relation to the CCR (see previous section, Example 2). More specifically,
\begin{theorem} Let $\mathcal{C}_{L,\Pi}$ be a semi-P-Lagrangian (by flux components in $P$) constitutive relation of order 1: $F^{i}_{\mu}=L_{,z^{\mu}_{i}},\ (\mu,i)\in P$ where $L\in C^{\infty}(J^{1}_{p}(\pi))$. Let $\hat C_{L,\Pi}$ be a CCR covering $\mathcal{C}$ with $p=L-\sum_{(\mu,i)\in P}z^{\mu}_{i}\partial_{z^{\mu}_{i}}L $.  Then the
 following statements are equivalent
\begin{enumerate}
\item For  a section $s\in \Gamma(\pi)$ and for all  $\xi \in \mathcal{X}(J^{1}_{p}(\pi))$
\[ j^{1}_{p}(s)^{*}i_{\xi}{\tilde d}\Theta_{{\hat
C_{L,\Pi\ -}}}=0.  \]
\item Section $s$ is the solution of the system of balance equations
\beq j^{1}(s)^{*}E_{L}=\sum_{i\vert (\mu,i)\in P}\left[ (L_{,z^{\mu}_{i}}\circ j^{1}_{p}(s))_{;i}-L_{,y^\mu}\circ j^{1}_{p}(s)\right] +\sum_{(\mu,i)\notin P}(F^{i}_{\mu})_{;i} =\Pi_{\mu}( j^{1}_{p}(s)).
\eeq
\end{enumerate}
\end{theorem}
\begin{remark} In a case of a Lagrangian Field Theory of higher order ($k>1$) the Euler-Lagrange system of equations does not have unique representation as a balance system. Perhaps this is the reason why the last results are limited to the Lagrangian system of the first order.  Probably introduction of more general form of balance systems allows to include higher order Lagrangian systems in this scheme in such a way that the arbitrary vector fields $ \xi$ in the $k$-jet bundles are admissible as variations.
\end{remark}
Similarly, there are no restriction to the variation in the RET case where CR is the section of the bundle $\pi^{n+(n+1)}:\Lambda^{n+(n+1)}_{2/1}Y$ (see \cite{Pr1}).
\begin{theorem} Let $\mathcal{C}$ be a constitutive relation of the RET type and $\tilde C$ - corresponding \emph{lifted} CCR. Then, for a given section $s\in \Gamma(\pi)$
\[ s^{*}(i_{\xi}{\tilde d}\Theta_{{\hat C}_{-}})= 0\ \forall\ \xi \in \mathcal{X}(Y) \Leftrightarrow \text{Section\ s\ is\  the\  solution\  of}\ \bigstar.\]
\end{theorem}

In general case we have to introduce the class of \emph{admissible variations}.

\begin{definition}
 A vector field $\xi \in \mathcal{X}(U),\ U\subset J^{k}_{p}(\pi)$ is called $C$-admissible ($\xi \in \mathcal{X_{C}}(U)$)  if
 \begin{enumerate}
 \item In the case $k=1$, for some (=any) prolongation of $\xi $ to the vector field $\xi^2 \in \mathcal{X}(J^{2}_{p}(\pi^{ -1}_{21}(U))$
    \[ \omega^{2}_{\tilde{C}}(\xi^2)=\sum_{(\mu,i)\in P}(\xi^{\mu}_{i}- \xi^{k}z^{\mu}_{i k})F^{i}_{\mu}=0. \]
    \item For $k>1,$ for some (=any) prolongation of $\xi $ to the vector field $\xi^{k+1} \in \mathcal{X}(J^{k+1}_{p}(\pi^{ -1}_{(k+1)1}(U))$
    \[
     \omega^{k+1}_{\widetilde{C}}(\xi)=\sum_{(\mu,i )\in P}(\xi^{\mu}_{i}- \xi^{k}z^{\mu}_{ik })F^{i}_{\mu}=0.
    \]
    \end{enumerate}
    \end{definition}
    \begin{remark} Condition of $C$-admissibility is much more restrictive in the case $k=1$ then in the case $k>1$ since in the last case $F^{i}_{k}$ may depend on $z^{\mu}_{ik}$.
    \end{remark}
As the next result shows, locally any CR is separable.
 \begin{lemma} Let $W\subset Y$ be the domain of a fiber chart $(x^i ,y^\mu)$.  Any CR $C$ is \textbf{separable} in $W^k=\pi_{k0}^{-1}(W)$, more specifically, $\forall y\in W,\ z\in W^k,\pi_{k0}(z)=y,$ the mapping $\mathcal{X}_{C}(W^k)\ni \xi \rightarrow \{ \omega^{\mu}_{z}(\xi) \} \in R^m$ is the epimorphism.
 \end{lemma}
 \begin{proof} Consider vertical vector fields that are constant along the fiber $U\cap W$.
 \end{proof}
 As a result, in general case we have the following result
\begin{theorem}
For any CR $\mathcal{C}$ and a domain $U\subset J^{k}_{p}(\pi)$ the following  statements
for a section $s\in \Gamma(\pi)(U),\ U\subset X$ are equivalent:
\begin{enumerate}
\item $\mathcal{C}$ is separable in $U$ and for all $\xi \in \Gamma
(U,\mathcal{V}_{\mathcal{C}}(J^{k}_{p}(\pi)),$
\[ j^{k}(s)^{*}(i_{\xi}{\tilde d}\Theta_{{\tilde C}_{-}})=0.\hskip7cm (IBS)\]
\item Section $s\in \Gamma(U,\pi)$ is the solution of the system
\[
 (F^{i}_{\mu}\circ j^{k}_{p}(s))_{,x^{i}}+F^{i}_{\mu}\circ j^{k}_{p}(s)\lambda_{G,i}=
 \Pi_{\mu}(j^{k}_{p}(s)),\ \mu=1,\ldots, m.\hskip1cm (\bigstar)
\]
\end{enumerate}
\end{theorem}

\subsection{Contact source correction of the lifted PC-form $\widetilde{\Theta}_{\hat C}$ and the balance equations.}
\vskip0.4cm
In this subsection we show that one may modify the Poincare-Cartan form $\Theta_{\tilde{C}}$ by adding a contact term to the source in such a way that applying the procedure of variation described above to the PC-form modified in this way one can remove the restriction s ont the variational vector fields.\par
Let $\Lambda^{n+(n+1)}_{2}(J^{1}_{p}(\pi))\rightarrow J^{1}_{p}(\pi)$ be the bundle of $n+(n+1)$-forms on the partial 1-jet bundle $J^{1}_{p}(\pi)$ annulated by substitution of any two $\pi^{1}$- vertical vector fields.  An element of a fiber of this bundle have the form
\beq
p\eta +p^{i}_{\mu}dy^\mu \wedge \eta_{i}+ p_{\mu} dy^\mu \wedge \eta + k^{i j}_{\mu}dz^{\mu}_{i}\wedge \eta_{j} + q^{i}_{\mu}dz^{\mu}_{i}\wedge \eta .\eeq
Introduce on this manifold the following $n+(n+1)$-form:
\beq
Q=p\eta +p^{i}_{\mu}dy^\mu \wedge \eta_{i}+ p_{\mu} dy^\mu \wedge \eta  + q^{i}_{\mu}dz^{\mu}_{i}\wedge \eta =
p\eta +p^{i}_{\mu}dy^\mu \wedge \eta_{i}+ p_{\mu} \omega^{\mu} \wedge \eta + q^{i}_{\mu}\omega^{\mu}_{i}\wedge \eta
\eeq
which differs from the canonical form $\Theta^{n}_{2}+\Theta^{n+1}_{2}$ by the last term only.  It is easy to see that this form behaves covariantly under the change of fibred coordinates.\par

Let $\hat C :J^{k}_{p}(\pi)\rightarrow \Lambda^{n+(n+1)}_{2}(Y)$ be a covering constitutive relation of the form $\hat C (x,y,z)=p\eta +F^{i}_{\mu}dy^\mu \wedge \eta_{i}- \Pi_{\mu} dy^\mu \wedge \eta $.  Mapping $\hat C$ can be lifted to the mapping
\[
\hat C^{1}: J^{k}_{p}(\pi)\rightarrow \Lambda^{n+(n+1)}_{2}(J^{1}_{p}(\pi)): (x,y,z)\rightarrow p\eta +F^{i}_{\mu}dy^\mu \wedge \eta_{i}-K_{\mathcal{C}} ,
\]
where the "source" form $K_\mathcal{C} =\Pi_{\mu} \omega^\mu \wedge \eta +F^{i}_{\mu}\omega^{\mu}_{i}\wedge \eta$ was introduced in the previous section.\par
Now we will modify the Poincare-Cartan form $\Theta_{\hat C}$ on the (partial) jet bundle $J^{k}_{p}(\pi)$ for $k>1$ and on the partial 2-jet bundle $J^{2}_{p}(\pi)$ for $k=1$ by adding an extra source term to get the modified covering Poincare-Cartan form $\Theta_{\hat C}$ by the formula
\beq
\widehat{\Theta}_{\hat C}=\hat C^{1\ *}Q=  p\eta +F^{i}_{\mu}dy^\mu \wedge \eta_{i}-\Pi_{\mu}\omega^\mu \wedge \eta -F^{i}_{\mu}\omega^{\mu}_{i}\wedge \eta.
\eeq
Applying to the form $\Theta_{\hat C}$ arguments leading to (5.4-5.7) we get, for a lift $\xi^2$ to the bundle $J^{2}_{p}(\pi)$ of a vector field $\xi \in \mathcal{X}(J^{1}_{p}(\pi))$ for $k=1$ and for a lift $\xi^{k+1}\in \mathcal{X}(J^{k+1}_{p}(\pi))$ of the vector field $\xi^{k}$ on $J^{k}_{p}(\pi)$ for $k>1$
\begin{multline}
i_{\xi^2}{\tilde d}\widehat{\Theta}_{\hat C}=\{-\omega^{i}(\xi)[d_{i}F^{i}_{\mu}+\lambda_{G,i}F^{i}_{\mu}-\Pi_{\mu}-\partial_{y^\mu}(p+z^{\nu}_{j}F^{j}_{\nu})]-
\omega^{\mu}_{i}(\xi^2)[ F^{i}_{\mu}-\partial_{z^{\mu}_{i}}(p+z^{\nu}_{j}F^{j}_{\nu})]\}\eta + \\ + F^{i}_{\mu}\omega^{\mu}_{i}(\xi^2)\eta -F^{i}_{\mu}\omega^{\mu}_{i}\wedge i_{\xi}\eta+ Con=\\ =
\{-\omega^{\mu}(\xi)[d_{i}F^{i}_{\mu}+\lambda_{G,i}F^{i}_{\mu}-\Pi_{\mu}-\partial_{y^\mu}(p+z^{\nu}_{j}F^{j}_{\nu})]+
\omega^{\mu}_{i}(\xi^2)[ \partial_{z^{\mu}_{i}}(p+z^{\nu}_{j}F^{j}_{\nu})]\}\eta + Con,
\end{multline}
since the second to the last last form in the previous expression is contact. For the \emph{lifted covering constitutive relation} $\tilde C$  (see Def.6) this expression simplifies:
\beq
i_{\xi^2}{\tilde d}\widehat{\Theta}_{\tilde C}=-\omega^{\mu}(\xi)[d_{i}F^{i}_{\mu}+\lambda_{G,i}F^{i}_{\mu}-\Pi_{\mu}]\eta +Con.
\eeq
Thus, we have proves the following
\begin{theorem} For a given constitutive relation $\mathcal{C}$ of order $k$ and a section $s\in \Gamma(\pi )$ of the bundle $\pi$ the following statements are equivalent:
\begin{enumerate}
\item  For any vector field $\xi \in \mathcal{X}(J^{k}_{p}(\pi))$ and its arbitrary prolongation to the $\pi^{k+1}_{k}$-projectable vector field $\xi^{k+1}$ on $J^{k+1}_{p}(\pi)$,
    \beq
    j^{k+1\ *}(s)i_{\xi^{k+1}}{\tilde d}\widehat{\Theta}_{\tilde C_{-}}=0,
    \eeq
    \item Section $s$ is the solution of the system $\bigstar$: $ \left( F^{i}_{\mu}\circ j^{k}(s)\right)_{;i}=\Pi_{\mu}\circ j^{k+1}(s),\ \mu=1,\ldots ,m$.
\end{enumerate}
\end{theorem}
\subsection{Dual form of a balance system}

Reversing in the IBS equation (5.1) the order of contraction with a vector field $\xi^k$ and acting by the Iglesias differential (but keeping the sign of the source term) we get the equation
\beq
j^{k+1}(s)^{*}\tilde{d}i_{\xi^k}\Theta_{\hat C}=0.
\eeq
Calculations similar to one performed above for the equation (5.1) allows to prove the following

\begin{proposition} Let $\hat C$ be a CCR and let $\Theta_{\hat C}=p\eta +F^{\mu}_{i}dy^i \wedge \eta_{\mu}+\Pi_{i}dy^i \wedge \eta$ be the corresponding Poincare-Cartan form.  Then, for any projectable vector field $\xi \in X_{p}(\pi)$,

\begin{multline}
\tilde{d}i_{\xi^k }\Theta_{\hat C}=[\omega^{\mu}(\xi)[d_{i}F^{i}_{\mu}+F^{i}_{\mu}\lambda_{G,i}-\Pi_{\mu}]+
\xi^{i}(d_{i}+\lambda_{G,i})\cdot (p+\sum_{(\mu,j)\in P }z^{\mu}_{j}F^{j}_{\mu})+\\ + (p+\sum_{(\mu,j)\in  P }z^{\mu}_{j}F^{j}_{\mu})d_{i}\xi^{i}+F^{i}_{\mu}d_{i}\omega^{\mu}(\xi)\eta .
\end{multline}
For the lifted CCR $\tilde C$, where $p=-\sum_{(\mu,i)\in P}z^{\mu}_{i}F^{i}_{\mu}$, the previous formula takes the form
\beq
\tilde{d}i_{\xi^k }\Theta_{\hat C}=\left[\omega^{\mu}(\xi)[d_{i}F^{i}_{\mu}+F^{i}_{\mu}\lambda_{G,i}-\Pi_{\mu}]+F^{i}_{\mu}d_{i}\omega^{\mu}(\xi)\right]\eta
\eeq
\end{proposition}
\begin{remark} Notice that the condition $F^{i}_{\mu}d_{i}\omega^{\mu}(\xi)=0$ is equivalent, for the vertical vector fields $\xi \in V(\pi)$ to the admissibility condition of Def.7.
\end{remark}
\begin{corollary} Let $C$ be a constitutive relation of order $k\geqq 1$ and let $\tilde C$ be the corresponding lifted CCR. Then for a section $s\in \Gamma(U,\pi),\ U\subset X$ defined in an open subset $U\subset X$ such that $C$ is separable in $\pi^{-1}(U)$, the following conditions are equivalent
\begin{enumerate}
\item For all $C$-admissible vector fields $\xi \in \pi^{-1}(U)$
\[
j^{k+1}(s)^{*}\tilde{d}i_{\xi^k}\Theta_{\hat C}=0.
\]

\item $s$ is the solution of the balance system $\bigstar$: \[ \left( F^{i}_{\mu}\circ j^{1}(s)\right)_{;i}=\Pi_{\mu}\circ j^{1}(s),\ \mu=1,\ldots ,m.\]
\end{enumerate}
\end{corollary}
\begin{remark} Expression in the left side of equation (5.17) applied to the $\Theta^{n}_{\tilde C}$ represents the second, boundary term in the Cartan formula for the Lie derivative
\[
\mathcal{L}_{\xi}\Theta^{n}_{\tilde C}=i_{\xi}d\Theta^{n}_{\tilde C}+di_{\xi}\Theta^{n}_{\tilde C}.
\]
It is interesting that one can use both these terms to get the balance system $\bigstar$.
\end{remark}

\section{Geometrical transformations and the symmetry groups of constitutive relations.}

 Let $\hat C$ be a CCR with the domain $J^{k}_{p}(\pi)$, let $\phi \in Aut_{p}(\pi )$ be an automorphism of the bundle $\pi$ preserving the partial structure.  Denote by $\phi^{k}$ the flow lift of $\phi$ to to the manifold $J^{k}_{p}(\pi)$ and by $\phi^{1*}$ - its canonical flow lift to the space $\Lambda^{n+(n+1)}_{2}Y$

   Define the action of $\phi$ on the CCR $\hat C$ by the "conjugation"
\beq
{\hat C}^{\phi}=\phi^{1*}\circ {\hat C}\circ {\phi}^{k\ -1}.
\eeq
  Similarly one defines the action of $\phi$ on a CR $C$, using the projection $\widetilde{\phi}^{1*}$ of $\phi^{1*}$ to $\Lambda^{n+(n+1)}_{2/1}Y$ instead of $\phi^{1*}$, see \cite{KMS,LDS}. Immediate check shows that ${\hat C}^{\phi\circ \psi}=({\hat C}^{\psi})^{\phi}$. \par Since  the transformation $\phi^{1*}$ preserves the canonical multisymplectic form on $\Lambda^{n+(n+1)}_{2}Y$ it is easy to see that for the CCR ${\hat C}^{\phi}$ one has
  \beq \Theta_{{\hat C}^\phi}=\phi^{k\ -1*}\Theta_{\hat C}.\eeq\par

\subsection{Geometrical symmetries of CR and CCR}
\begin{definition}
\begin{enumerate}
\item
An automorphism $\phi \in Aut_{p}(\pi)$ is called a \textbf{ geometrical symmetry transformation} of a CR
$\mathcal{C}$ (resp. of a CCR $\widehat{\mathcal{C}}$) if
\[\mathcal{C}^{\phi}=\mathcal{C}\ \  (\text{respectively},\  \widehat{\mathcal{C}}^{\phi}=\widehat{\mathcal{C}}).\]
\item
Let $\xi\in \mathcal{X}_{p}(\pi)$ be a $\pi$-projectable vector field.  $\xi $ is called a
\textbf{geometrical infinitesimal symmetry} of the CR $\mathcal{C}$ (resp. of a CCR $\widehat{\mathcal{C}}$) if (see \cite{KMS}, or below for the definition of a Lie derivative along the mapping)
\[ L_{(\xi^{k},\widetilde{\xi}^{1*})}\mathcal{C}=0,\]
(respectively, $L_{(\xi^{k},{\xi}^{1*})}\widehat{\mathcal{C}}=0$).
\end{enumerate}
\end{definition}
In the next proposition we collected some simple properties of geometrical symmetries and infinitesimal symmetries of constitutive relations and covering constitutive relations.  These properties follows directly from definitions and the fact that the transformation $\phi^{1*}$ preserves the canonical multisymplectic form on $\Lambda^{n+(n+1)}_{2}Y$. In this proposition we put
\[
W_{p}=J^{k}_{p}(\pi)\times \Lambda^{n+(n+1)}_{2/1}(Y), \hat{W}_{p}=J^{k}_{p}(\pi)\times \Lambda^{n+(n+1)}_{2}(Y)
\]
\begin{proposition}  For a CR $\mathcal C$ (respectively for a CCR $\hat C$)
\begin{enumerate}
\item Geometrical symmetries $\phi$ of $C$ (resp. $\hat C$) form the subgroup $Sym(\mathcal{C})\subset Aut_{p}(\pi)$.
\item Infinitesimal symmetries of $\mathcal{C}$ (resp. $\hat C$) form the Lie algebra $\mathfrak{g}(\mathcal{C})\subset \mathcal{X}_{p}(\pi)$
with the bracket of vector  fields in $Y$ as the Lie algebra operation.
\item A vector field $X\in \mathcal{X}(W_{p})$ (resp. $X\in \mathcal{X}(\hat{W}_{p})$) is the generator of the 1-parametrical group of generalized
 symmetries of $\mathcal{C}$ (resp. $\hat C$) if and only if it is tangent to the graph $\Gamma_{\mathcal{C}}$  (resp.  $\Gamma_{\widehat{\mathcal{C}}}$).

 \item A vector field $\xi \in \mathcal{X}_{p}(\pi)$ is an infinitesimal symmetry of $\mathcal{C}$ if
 (and only if) the (local) phase flow diffeomorphisms $\Phi^{\xi}_{t}=\psi_{t}\times \phi^{k}_{-t}$ of $W_{p}$ (resp. of $\hat{W}_{p}$) defined by the prolongation of $\xi$ maps $\Gamma_{\mathcal{C}}$ into itself satisfies to the relation
\[
\Phi^{\xi}_{t}(\Gamma_{\mathcal{C}})=\Gamma_{\mathcal{C}}\ \text{resp.}\ ,\Phi^{\xi}_{t}(\Gamma_{\widehat{\mathcal{C}}})=\Gamma_{\widehat{\mathcal{C}}} .
\]
i.e. if the (local) phase flow $\phi_{t}$ of vector field $\xi$ is formed by the geometrical symmetry transformations of $\mathcal{C}$ (resp. of $\widehat{\mathcal{C}}$ ).

\item Let $\xi $ be an infinitesimal symmetry of a CR $C$ (respectively of a covering constitutive relation $\hat C$). Then, for the local phase flow $\phi_{t}$ of $\xi$ one has
\[\Theta_{{ C}^{\phi_{t}}}=\phi^{k\ -1\ *}_{t}\Theta_{ C}=\Theta_{ C}\Leftrightarrow \mathcal{L}_{\xi^l}\Theta_{C}=0,\]
(respectively, $\mathcal{L}_{\xi^k}\Theta_{\hat{C}}=0$).
\end{enumerate}
\end{proposition}

In the next statement we collect the basic properties of action of automorphisms $\phi \in Aut_{p}(\pi )$, including geometrical symmetries of constitutive relations, on the balance systems $B_{C}$. For a domain $U\subset X$ denote by $Sol_{\mathcal{C}}(U)$ the space of solutions $s:U\rightarrow Y$ of the balance system $\bigstar$. correspondence $U\rightarrow Sol(U)$ determine the sub-sheaf on the manifold $X$. Denote by $Sol_{\mathcal{C}}$ the sheaf corresponding to the introduced subsheaf.
\begin{theorem}
\begin{enumerate}
\item Mapping $\xi \rightarrow \phi_{*}\xi$ maps the space of $\mathcal{C}$-admissible vector fields
$\mathcal{X}_{\mathcal{C}}(W)$ at a domain $W$ onto the space of $\mathcal{C}^\phi$-admissible vector fields $X_{\mathcal{C}^\phi
}(\phi(W))$ in the domain $\phi(W)$ for any open subset $W\subset J^{k}_{p}(\pi)$. This mapping defines the
isomorphism of (pre)-sheaves of admissible vector fields $\mathcal{X}_{C}\rightleftarrows
\mathcal{X}_{\mathcal{C}^\phi }$.
\item Mapping $s \rightarrow s^\phi= \phi \circ s \circ {\bar \phi}^{-1}$, on the sections of the
configurational bundle $\pi:Y\rightarrow X$ maps the space of solutions $Sol_\mathcal{C}(W)$ in the domain $W\subset X$ onto the space of
solutions $Sol_\mathcal{C^\phi}(\phi(W))$ of the balance system $\mathcal{B_{C^\phi}}$ in the domain $\phi(W)\subset X$:
\[  Sol_{C}(W) \rightleftarrows  Sol_{C^\phi }(\phi(W)). \]
\item Let $\phi \in Aut_{p}(\pi)$ be a symmetry of the CR
 $\mathcal{C}$. Then the mapping $s\rightarrow
 s^\phi =$ maps the sheaf $Sol_C$ of solutions of the balance system $\bigstar $ into itself.
\end{enumerate}
\end{theorem}
Proof of this Theorem is based on the following statement

\begin{lemma} Let $\hat C$ be a CCR with the domain $J_{p}^{k}(\pi)$ and the Poincare-Cartan form
$\Theta_{\hat C}=p\eta +F^{\mu}_{i}dy^i \wedge \eta_{\mu}+\Pi_{i}dy^i \wedge \eta$ and let $\widehat{\phi}
\in Aut_{p}(\pi^{k})$ be an automorphism of the double bundle $J_{p}^{k}(\pi)\rightarrow Y\rightarrow X$.  Then

\beq [\widehat{\phi}^{*}\Theta_{\hat{C}}]=^{\widehat{\phi}}p \eta + ^{\widehat{\phi}}F^{\mu}_{i}dy^i
\wedge \eta_{\mu}+ ^{\widehat{\phi}}\Pi_{i}dy^i \wedge \eta, \eeq where
 \beq
\begin{cases}
^{\widehat{\phi}}p= [p\circ \widehat{\phi}+ (F^{\mu}_{i}\circ \widehat{\phi})(\phi^{i}_{,x^\nu}J({\bar
\phi}^{-1})^{\nu}_{\mu}]\cdot detJ({\bar \phi}),
 \\  ^{\widehat{\phi}}F^{\mu}_{i}= detJ({\bar \phi})J({\bar
\phi}^{-1})^{\mu}_{\nu}(F^{\nu}_{j}\circ
\widehat{\phi})\phi^{j}_{,y^i},\ \\
^{\widehat{\phi}}\Pi_{i}=detJ({\bar \phi})(\Pi_{j}\circ \widehat{\phi}) \phi^{j}_{,y^i}.
\end{cases}
\eeq
\end{lemma}
\begin{proof}(of Lemma).
We notice that ${\bar \phi}^{*}\eta =(detJ({\bar \phi}))\eta$ where $detJ({\bar
\phi})$ is the Jacobian of the (local) diffeomorphism $\bar \phi$ defined by the volume form $\eta$.  On
the other hand
\[
{\bar \phi}^{*}\eta_{\mu}={\bar \phi}^{*}i_{\partial_{x^\mu}}\eta=i_{{{\bar
\phi}^{-1}}_{*}\partial_{x^\mu}}{\bar \phi}^{*}\eta=detJ({\bar \phi})J({\bar
\phi}^{-1})^{\nu}_{\mu}\eta_{\nu}
\]
since ${{\bar \phi}^{-1}}_{*}\partial_{x^\mu}= J({\bar \phi}^{-1})^{\nu}_{\mu}\partial_{x^\nu}.$\par
Altogether we have
\begin{multline}
\widehat{\phi}^{*}\Theta_{\hat{C}}= \widehat{\phi}^{*}[p\eta+F^{\mu}_{i}dy^i \wedge \eta_{\mu}+\Pi_{i}dy^i
\wedge \eta]=(p\circ \widehat{\phi})\cdot {\bar \phi}^{*}\eta +\\ + (F^{\mu}_{i}\circ
\widehat{\phi})d(\phi^{ i})(x,y)\wedge {\bar \phi}^{*}\eta_{\mu}+\Pi_{i}\circ \widehat{\phi} d(\phi^{ i})
\wedge {\bar \phi}^{ *}\eta=\\
=p\circ \widehat{\phi}\cdot detJ({\bar \phi})\eta +(F^{\mu}_{i}\circ
\widehat{\phi})(\phi^{i}_{,x^\sigma}dx^\sigma +\phi^{i}_{,y^j}dy^j)\wedge [detJ({\bar \phi})J({\bar
\phi}^{-1})^{\nu}_{\mu}\eta_{\nu}]+\\ +\Pi_{i}\circ \widehat{\phi}(\phi^{i}_{,x^\sigma}dx^\sigma
+\phi^{i}_{,y^j}dy^j)\wedge detJ({\bar \phi})\eta = [detJ({\bar \phi})J({\bar
\phi}^{-1})^{\nu}_{\mu}(F^{\mu}_{i}\circ \widehat{\phi})\phi^{i}_{,y^j}] dy^j \wedge \eta_{\nu}+\\ +
[detJ({\bar \phi})(\Pi_{i}\circ \widehat{\phi}) \phi^{i}_{,y^j}]dy^j \wedge \eta +[p\circ
\widehat{\phi}\cdot detJ({\bar \phi})+ (F^{\mu}_{i}\circ \widehat{\phi})[\phi^{i}_{,x^\nu}detJ({\bar
\phi})J({\bar \phi}^{-1})^{\nu}_{\mu}]\eta.
\end{multline}
Splitting the terms we get the result stated in Lemma.
\end{proof}

\begin{proof}\textbf{Proof of Theorem 7.}
Transformation $\phi$, generating the transformation of a constitutive relations $C\rightarrow
\mathcal{C}^\phi $, transforms the (1,n)-component $F^{\mu}_{i}$ of $C$ into the corresponding component $
F^{\phi \ \mu}_{i}= detJ({\bar \phi}^{-1})J({\bar \phi})^{\mu}_{\nu}(F^{\nu}_{j}\circ
(\phi^1)^{-1})\phi^{-1\ j}_{,y^i}$ of the CR $C^\phi$, see last Lemma.  On the other hand the mapping $\phi_{*}$ maps a
vector field $\hat \xi=\xi^{\mu}\partial_{x^\mu}+\xi^{i}\partial_{y^i}+\xi^{i}_{\mu}\partial_{z^{i}_{\mu}}+\ldots
\in \mathcal{X }(J^{k}_{p}(\pi))$ into the vector field with the local $z^{i}_{\mu}$-component having the form
\[
\phi_{*}\xi = \ldots   +\phi^{ j}_{,y^i} J({\bar \phi}^{-1})^{\mu}_{\nu}\xi^{i}_{\mu}\partial_{z^{j}_{\nu}}+\ldots ,
\]
see Appendix II in \cite{Pr1}.  As a result for the value of the pairing  we have

\[ F^{\phi\ \mu}_{i}(\phi_{*}\xi)^{i}_{\mu}=\left( F^{ \mu}_{i}\xi^{i}_{\mu}\circ \phi^1 \right) \cdot detJ({\bar \phi}^{-1}). \]
Since $detJ({\bar \phi}^{-1})>0$, expressions for contraction of flux components and components of lifted vector field in left and right sides vanish simultaneously.  This proved the first statement in Theorem 6.\par

To prove the second statement notice that
\begin{multline}
(j^{1}_{p}(s^\phi ))^{*}i_{\phi_{*}\xi}{\tilde d}\Theta_{\widehat{\mathcal{C}}^\phi_{-}}=[\phi^{1\
*}(j^{1}_{p}(s))]^{*}i_{\phi_{*}\xi}{\tilde d}\Theta_{\widehat{\mathcal{C}}^\phi_{-}}=
(j^{1}_{p}(s))^{*}\circ \phi^{1\  *}i_{\phi_{*}\xi}{\tilde d}\Theta_{\widehat{\mathcal{C}}^\phi_{-}}=\\=
(j^{1}_{p}(s))^{*}\circ i_{\phi^{1\ -1}_{*}\phi_{*}\xi}{\tilde d}\phi^{1\
*}\Theta_{\widehat{\mathcal{C}}^\phi _{-}}=(j^{1}_{p}(s))^{*}\circ i_{\xi}{\tilde
d}\Theta_{(\widehat{\mathcal{C}}^\phi )^{\phi^{-1}}_{-}}=(j^{1}_{p}(s))^{*}\circ i_{\xi}{\tilde
d}\Theta_{\widehat{\mathcal{C }}_{-}}.
\end{multline}
Since the mapping $\xi \rightarrow \phi_{*}\xi$ is, by the first statement the isomorphism of the
pre-sheaves of admissible vector fields,  expression in the right side is zero for all $\xi \in \mathcal{X}_{\mathcal{C}}(W)$
if and only if the expression in the left side is zero for all vector fields $\xi \in \mathcal{X}_{\mathcal{C}^\phi
}(\phi(W))$ and, therefore $s \in Sol_{\mathcal{C}}(W)$ if and only if $s^\phi \in Sol_{\mathcal{C}^\phi }(\phi (W)).$\par

For the proof of the third statement we use the symmetry condition in the form presented at the end of Definition 8
and notice that in this case $\mathcal{C}^\phi =\mathcal{C}$ so the mapping $s\rightarrow s^\phi$ maps the space $Sol_{\mathcal{C}}(W)$
 isomorphically onto the space $Sol_ {\mathcal{C}}(\phi (W))$.
\end{proof}

\subsection{Infinitesimal symmetries of the constitutive relations.}
Here we consider the infinitesimal transformations of a CCR $\hat C$ and these of a CR $\mathcal C$ of the first order.  We start with an arbitrary  covering constitutive relation ${\hat C}:z=(x^\mu,y^i,z^{i}_{\mu})\rightarrow (p(z),F^{\mu}_{i}(z),\Pi_{k}(z))$.\par
Let
\[ {\hat \xi}=\xi^{\mu}(x)\partial_{x^\mu}+\xi^{i}(x,y)\partial_{y^i}+\xi^{i}_{\mu}(x,y,z)\partial_{z^{i}_{\mu}}=\xi+ \xi^{i}_{\mu}(x,y,z)\partial_{z^{i}_{\mu}}
\]
be an arbitrary infinitesimal automorphism of the double bundle $J^{1}_{p}(\pi)\rightarrow Y\rightarrow X$.  Here we introduced notation $\xi$ for the projection of $\hat \xi$ to $Y$.\par

In particular, let $\xi=\xi^{\mu}(x)\partial_{x^\mu}+\xi^{i}(x,y)\partial_{y^i} \in \mathcal{X}_{p}(\pi)$  be an \emph{infinitesimal automorphism} (vector field) of the
bundle $\pi$, i.e. a projectable vector field in $Y$ satisfying to the conditions of Sec.3.3 for lifting
to the partial 1-jet bundle $J^{1}_{p}(\pi).$  Let, as before, $\xi^{1}$ be its prolongation to the projectable contact vector field in $J^{1}_{p}(\pi)$.
Then, as an example of vector fields $\hat \xi$ one may consider the flow lift (see (3.2))
\[
\xi^{1}=\xi^{\mu}(x)\partial_{x^{\mu}}+\xi^{i}(x,y)\partial_{y^{i}} +\left(d_{\mu}\xi^{i}
-z^{i}_{\nu}\frac{\partial \xi^{\nu}}{\partial x^{\mu}} \right)\partial_{z^{i}_{\mu}},
\]
of $\xi$ to $J^{1}_{p}(\pi)$. Here summation in the last term is taken over the $z^{i}_{\mu}$ that are present in the partial 1-jet
bundle. In the RET case we do not need to introduce any prolongation.
\par

Projection $\xi$ of a vector field ${\hat \xi}$ defines the
prolongation ${\xi}^{1*}$ to the projectable vector field in $ Z*=\Lambda^{(n+1)+(n+2)}_{2}Y$
preserving canonical multisymplectic form(s) $\Theta^{n}_{2}+\Theta^{n+1}_{2}$ (See Sec.7) and, the prolongation $\hat \xi^{1*}$ to the bundle $\Lambda^{(n+1)+(n+2)}_{2}(J^{1}(\pi ))$:
\begin{multline}
{\xi}^{1*}=\xi^{\mu}(x)\partial_{x^{\mu}}+\xi^{i}(x,y)\partial_{y^{i}}+
\left( -p\cdot div_{G}({\bar \xi}) -p^{\mu}_{i}\frac{\partial \xi^{i}}{\partial x^\mu }   \right) \partial_{p}+\\
+\left(p^{\nu}_{i}\frac{\partial \xi^\mu }{\partial x^\nu
}-p^{\mu}_{j}\frac{\partial \xi^j }{\partial y^i}-p^{\mu}_{i}div_{G}({\bar \xi}) \right)
\partial_{p^{\mu}_{i}}+\left( -q_{k}div_{G}({\bar \xi})-q_{j}\frac{\partial \xi^j}{\partial y^k} -q^{\mu}_{i}\xi^{i}_{\mu,y^k}\right)
\partial_{q_k}+\\+ \left(-q^{\mu}_{i}\xi^{i}_{\mu,z^{j}_{\nu}}-q^{\nu}_{j}div_{G}(\bar \xi)\right)\partial_{q^{\nu}_{j}}.
\end{multline}
We have used here the equality $ \xi^{i}_{,z^{j}_{\nu}}=0$ valid for automorphisms of the double bundle $J^{1}_{p}(\pi)\rightarrow Y\rightarrow X$. It is easy to see that vector field ${\xi}^{1*}$ is projectable to $\Lambda^{(n+1)+(n+2)}_{2/1}J^{1}_{p}(\pi).$  This projection - $\widetilde{{\xi}^{1*}}$ has, in fibred chart, expression (6.7) with the third term omitted.\par

Let now $\widehat{\phi}_{t}$ be a local flow in $J^{1}_{p}(\pi)$ of the vector field $\hat{\xi}$ and $\psi_{t}$ be a
local flow in $ \Lambda^{(n+1)+(n+2)}_{2}(J^{1}(\pi))$ of the vector field ${\xi}^{1*}$ (respectively $\widetilde{\psi}_{t}$ be a
local flow in $ \Lambda^{(n+1)+(n+2)}_{2/1}Y$ of the vector field $\widetilde{{\xi}}^{1*}$). \par

 Derivating by $t$ at $t=0$ the expression for the transformed mapping $\psi_{t} \circ \hat{C}\circ \hat{\phi}_{-t}$ (respectively for $\psi_{t} \circ \hat{C}\circ \hat{ \phi}_{-t}$)  we get the \emph{generalized Lie derivative} of mapping $\widehat{C}$ with respect to the vector fields
$(\hat{\xi},{\xi}^{1*})$ (see \cite{KMS}, Chapter 11) - the vector field over the mapping
$\hat{C}:J^{1}_{p}(\pi)\rightarrow \Lambda^{(n+1)+(n+2)}_{2}Y$ (respectively $C:J^{1}_{p}(\pi)\rightarrow {\tilde Z}$ for CR $\mathcal C$):
\beq \mathcal{L}_{(\hat{\xi},\widetilde{\xi}^{*})}\widehat{\mathcal{C}} =  \widetilde{\xi}^{*}\circ
\widehat{\mathcal{C}}-C_{*}(\hat{\xi}).\eeq
In local fibred coordinates we have for $\mathcal{L}_{(\hat{\xi},\widetilde{\xi}^{*})}\widehat{\mathcal{C}}$ the following expression
\begin{multline} \mathcal{L}_{(\hat{\xi},\xi^{1*})}\widehat{\mathcal{C}}=
-\left[ \hat{\xi}\cdot p+ div_{G}({\bar \xi})\cdot p  +F^{\mu}_{i}\frac{\partial \xi^{i}}{\partial x^\mu }      \right] \partial_{p}
-\left[ \hat{\xi}\cdot F^{\mu}_{k}+div_{G}({\bar \xi})F^{\mu}_{k}+F^{\mu}_{j}\frac{\partial \xi^j }{\partial y^k}- F^{\nu}_{k}\frac{\partial \xi^\mu }{\partial x^\nu} \right]
\partial_{p^{\mu}_{k}}-\\ - \left[ \hat{\xi}\cdot \Pi_{k}+div_{G}({\bar \xi})\Pi_{k}+\Pi_{j}\frac{\partial \xi^j}{\partial y^k}+F^{\mu}_{i}\xi^{i}_{\mu,y^k} \right]\partial_{q_{k}} - \left[\hat{\xi}\cdot F^{\nu}_{j}+F^{\nu}_{j}div_{G}(\bar \xi)+F^{\mu}_{i}\xi^{i}_{\mu,z^{j}_{\nu}}\right]\partial_{q^{\nu}_{j}},
\end{multline}
for a CR $\mathcal C$ the term with $\partial_{p}$ is absent from the expression of $L_{(\hat{\xi},\widetilde{\xi}^{1*})}\mathcal{C}$.\par

Condition that the generalized Lie bracket (6.9) is zero for a CCR $\widehat{\mathcal{C}}$ and a vector field $\xi \in X(Y)$, has the form of a system of differential equation of the first order for the components of the covering constitutive relation $\widehat{\mathcal{C}}$:
\beq
\begin{cases}
 \hat{\xi}\cdot p + div_{G}({\bar \xi}) p  +F^{\mu}_{i}\frac{\partial \xi^{i}}{\partial x^\mu }     =0,\\
  \hat{\xi}\cdot F^{\mu}_{k}+div_{G}({\bar \xi})F^{\mu}_{k}+F^{\mu}_{j}\frac{\partial \xi^j }{\partial y^k}- F^{\nu}_{k}\frac{\partial \xi^\mu }{\partial x^\nu
}  =0, \forall \mu,k,\\
 \hat{\xi}\cdot \Pi_{k}+div_{G}({\bar \xi})\Pi_{k}+\Pi_{j}\frac{\partial \xi^j}{\partial y^k}+F^{\mu}_{i}\xi^{i}_{\mu,y^k}=0, \forall k=1,\ldots, m,\\
 \hat{\xi}\cdot F^{\mu}_{k}+F^{\mu}_{k}div_{G}(\bar \xi)+F^{\nu}_{i}\xi^{i}_{\nu,z^{k}_{\mu}}=0,\ \forall k,\mu.
\end{cases}
\eeq

Vector field $\hat{\xi} = \xi^{\mu}{\partial_{\mu}}+\xi^{i}{\partial_i}+\xi^{i}_{\mu}\partial_{z^{i}_{\mu}}$ in these equations acts on the components of the
vector functions in the space $J^{1}_{p}(\pi)$. \par
If we subtract forth equation of this system from the second one we get
\beq
F^{\mu}_{j}\frac{\partial \xi^j }{\partial y^k}- F^{\nu}_{k}\frac{\partial \xi^\mu }{\partial x^\nu }-F^{\nu}_{i}\xi^{i}_{\nu,z^{k}_{\mu}}=0.
\eeq
This equation may replace the forth equation in the previous system. Yet, for $\hat \xi =\xi^1,\ \xi \in \mathcal{X}_{\pi}(Y)$ this equation is identically fulfilled.  To see this we remind (see ()) that in this case $\xi^{j}_{\nu}=d_\nu \xi^i-z^{j}_{\mu}\xi^{\mu}_{,\nu}=\xi^{j}_{,\nu}+z^{i}_{\nu}\xi^{j}_{,y^i}-z^{j}_{\sigma}\xi^{\sigma}_{,\nu}$ and, as a result, $\xi^{i}_{\nu,z^{k}_{\mu}}=\delta^{i}_{k}\delta^{\mu}_{\nu}\xi^{j}_{,y^i}-\delta^{j}_{k}\delta^{\mu}_{\sigma}\xi^{\sigma}_{,\nu}=
\delta^{\mu}_{\nu}\xi^{j}_{,y^k}-\delta^{j}_{k}\xi^{\mu}_{,\nu}.$  Substituting this in the equation () we will see that it is fulfilled identically.\par
Thus, for the case when $\hat \xi =\xi^1$ for a vector field $\xi \in \mathcal{X}_{p}(\pi)$ the use of modified CCR instead of the usual one does not add forth condition to the first three in the system (30.6) although it adds an extra term to the third equation.\par

We can rewrite this system of equations as the system of conditions \emph{for a vector field} $\xi \in \mathcal{X}_{p}(\pi )$ to be an infinitesimal symmetry of the CCR $\hat C$:
\beq
\begin{cases}
I:\ [p_{,z^{i}_{\mu}}\xi^{i}_{\mu}+F^{\mu}_{i}\xi^{i}_{,\mu}+p_{,i}\xi^{i}]+[pdiv_{G}({\bar \xi})+p_{,\mu}\xi^{\mu}]=0,\\
II^{\mu}_{k}:\  [F^{\mu}_{k,z^{i}_{\nu}}\xi^{i}_{\nu}+
F^{\mu}_{i}\xi^{i}_{,y^k}+F^{\mu}_{k,i}\xi^i] +
[F^{\mu}_{k}div_{G}({\bar \xi})-F^{\nu}_{k}\xi^{\mu}_{,\nu}
+F^{\mu}_{k,\nu}\xi^{\nu}]=0,\\
III_{k}:\ [\Pi_{k,z^{i}_{\mu}}\xi^{i}_{\mu}+\Pi_{i}\xi^{i}_{y^k}+ \Pi_{k,y^i }\xi^{i}]+[div_{G}({\bar \xi})\Pi_{k}+\Pi_{k,\mu}\xi^{\mu}]+F^{\mu}_{i}\xi^{i}_{\mu,y^k}=0.
\end{cases}
\eeq
In the case of a \emph{modified} CCR (see Sec.5.2) last condition contains both flux and source terms while for the original CCR last equation contains only source. Similar conditions an infinitesimal symmetry of a  CR $ \mathcal{C}$ are obtained from these by removing the first subsystem of equations.\par

Consider now the form these conditions takes in two special cases of vector fields:
\begin{example} $\pi$-\textbf{vertical vector fields}.\par
For a \textbf{vertical vector field} $\xi =\xi^{i}\partial_{y^{i}}\in \mathcal{V}(\pi)$ (infinitesimal gauge transformation) the system (12.18) takes the following form
\beq
\begin{cases}
I:\  p_{,z^{i}_\nu} d_{\nu}\xi^{i}+F^{\mu}_{i}\xi^{i}_{,\mu}+p_{,y^i}\xi^{i}    =0  ,\\
II^{\mu}_{k}:\  \left( F^{\mu}_{k,z^{j}_{\sigma}}d_{\sigma}+F^{\mu}_{j}\frac{\partial }{\partial
y^k}\right)\xi^{j} +F^{\mu}_{k,y^j}\xi^{j}=0,\ \mu=0,\ldots ,n;k=1,\ldots,m,\\
III_{k}:\  \left( \Pi_{k,z^{j}_{\sigma}}d_{\sigma}+\Pi_{j}\frac{\partial }{\partial
y^k}\right)\xi^j+\Pi_{k,y^j}\xi^{j}+F^{\mu}_{i}\xi^{i}_{\mu,y^k}=0,\ k=1,\ldots ,m.
\end{cases}
\eeq
\end{example}
\begin{example} $\nu$-\textbf{lifted vector fields}, $\nu$-trivial.\par
 Consider now mathematically simple but practically important case of a trivial bundle $\pi: Y=X\times U\rightarrow X$, fix a fibred chart $(W;x^\mu ,y^i)$ and consider the zero connection $\nu_{0}$ in this bundle defined in the domain $W\subset Y$ of the chart above by $\Gamma^{i}_{\mu}=0$. The $\nu_{0}$-horizontal lift of a basic vector field $\partial_\mu$ is $\hat{\partial}_{\mu}=\partial_\mu$ with zero vertical components. Its flow prolongation to $J^{i}_{p}(\pi)$ has the same form: $\hat{\partial}_{\mu}^{1}=\partial_\mu$. \par
Conditions (6.12) for the vector field $\hat{\partial}_{\mu}$ to define geometrical symmetry of a CCR $\hat C$ takes the form
\beq
\begin{cases}
I:\ p_{,\mu}+\lambda_{G,\mu}p=0,\\
II^{\nu}_{k}:\ F^{\nu}_{k,\mu}+ \lambda_{G,\mu}F^{\nu}_{k}=0,\\
III_{k}:\ \Pi_{k,\mu}+\lambda_{G,\mu}\Pi_{k}=0.
\end{cases}
\eeq
since $div_{G}(\partial_\mu )=\lambda_{G,\mu}$. These conditions require that the \emph{densities} of fluxes, source components and $p$ are invariant under the translation along $x^\mu$-axis. In a case where $\vert G\vert$-const these conditions reduces to the simple independence of all components of CCR $\hat C$ on the variable $x^\mu$.\par
\end{example}
Next result establish relation between the infinitesimal symmetries of a CCR and those of the corresponding CR.
\begin{proposition}
\begin{enumerate}
\item Any infinitesimal symmetry $\xi$ (respectively geometrical symmetry) of a CCR $\hat C$ generate (by the projection $\Lambda^{(n+1)+(n+2)}_{2}\rightarrow \Lambda^{(n+1)+(n+2)}_{2/1}$ the infinitesimal symmetry (respectively geometrical symmetry) of $\mathcal{C}$.
\item Let $\tilde C$ be the \textbf{lifted} CCR of a CR $\mathcal{C}$. Then the previous mapping defines the \textbf{bijection} between infinitesimal symmetries (generalized symmetries $\hat \xi $) of $C$ and $\tilde C$.
\end{enumerate}
\end{proposition}
\begin{proof}
First statement follows from the fact that second and third subsystems of the conditions (6.12) do not depend on $p$.\par
Second statement follows from the fact that for the lifted CCR, where $p=-z^{i}_{\mu}F^{\mu}_{i}$, multiplying equation $II^{\mu}_{k}$ by $z^{k}_{\mu}$ and sum by $k,\mu$ we get exactly equation I.
\end{proof}

\subsection{Homogeneous constitutive relations.}
If the state space of a theory contains enough background or dynamical fields to make the constitutive relation $\mathcal{C}$ free from the
explicit dependence of $\mathcal{C}$ on $(t,x)\in X$ (general relativity, theory of uniform materials and RET (see \cite{MR}) are three examples),
then the corresponding balance system simplifies and while studying it one does not need to introduce
assumptions on the character of the dependence of the balance system on a space-time point.  Definition given below
is an invariant way to distinguish a class of such CR.\par

Any local chart $x^\mu$ in $X$ defined the local (translational) action of $R^n$ in $X$ associating with
the basic vectors $e_{\mu}$ the vector fields $\partial_{x^\mu}.$  Vice versa, any n-dimensional
commutative subalgebra $\mathfrak{h}$ of the Lie algebra of vector fields $\mathcal{X}(U)$, $U$ being an
open connected subset of $X$, defines the locally transitive action of $R^n$ in $U$ and, therefore, a
local chart in a neighborhood of any point in $U$.

\begin{proposition}- \textbf{Definition}.  Let $\nu $ be a connection in the bundle $\pi$.   We will call a constitutive relation $\mathcal{C}$
\textbf{$\nu $-homogeneous} if the following \textbf{equivalent} properties of the constitutive relation $C$ are valid:
\begin{enumerate}
\item For any point $z\in J^{1}_{p}(\pi)$ there exists a local chart in a neighborhood
$U_{x},\ x=\pi^{1}(z)$ such that the (n+1)-Poincare-Cartan form $\Omega_{\mathcal{C}}$ of the CR $\mathcal{C}$
is invariant under the local flows $\phi^{\xi^1}_{t}$ of the lifts ${\hat \xi}^1$ of $\nu$-horizontal
vector fields ${\hat \xi}, \xi \in \mathfrak{h}$ in the neighborhood of $y=\pi_{10}(z)$:
\[
\mathcal{L}_{{\hat \xi}^1}\Omega_{\mathcal{C}}=0\ mod\ \mathbf{Con}.
\]
\item For all $\xi \in \mathfrak{h}$, the $\nu$-horizontal lift $\hat \xi$ is the infinitesimal symmetry
 of the constitutive mapping $\mathcal{C}$ in sense of Definition 8.
 \item The graph $\Gamma_{C}\subset J^{1}_{p}\times \tilde Z$ of mapping $\mathcal{C}$ is invariant
under the flow generated by (flow) lifts of $\nu$-horizontal vector fields $\hat \xi $, $\xi \in
\mathfrak{h}$.
\end{enumerate}
\end{proposition}
\begin{proof} Equivalence of statements of this proposition follows directly from the Proposition 11.
\end{proof}
 Similar definition and results formulated in the last Proposition can be formulated and proved for a CCR $\hat C$.
\begin{remark} In a case where connection $\nu$ is flat, the association $\xi \rightarrow {\hat \xi}^{1}$
defines the Lie algebra homomorphism $\mathfrak{h} \rightarrow Aut(\pi^{1})\subset \mathcal{X}(J^{1}_{p})$.
\end{remark}
Using first the $\nu$-horizontal lift of vector fields from $X$ to $Y$ and then the flow lift to $J^{1}_{p}(\pi)$ (Sec.3) one can get the $\nu$-homogeneity conditions of a CCR $\hat C$, those of lifted CCR $\tilde C$ or those of the CR $C$.

\section{Noether Theorem.}
In this section we present the (first) Noether theorem for balance systems corresponding to the action of a Lie group $G\subset Sym(C)\subset Aut_{p}(\pi)$ of the geometrical symmetries of a CR $\mathcal{C}$. We denote by  $\mathfrak{g}$ - the Lie algebra of the group $G$,  $\mathfrak{g}^{*}$ - its dual space, $\xi \rightarrow \xi_{Y}\in \mathcal{X}(Y)$ - action mapping on the manifold $Y$ and $\xi^k$ be its lift to the partial k-jet bundle $J^{k}_{p}(\pi)$, see Sec. 3.4.  \par
\subsection{Momentum Mapping and general Noether Theorem}
Condition that the group $G$ is the symmetry group of the CR $\mathcal{C}$ (and its lifted CCR $\tilde C$) has the infinitesimal form (see previous section)
\beq \mathcal{L}_{\xi^{k}_{Y}}(\Theta^{n}_{\tilde
C}+\Theta^{n+1}_{\tilde C})=0. \eeq
This condition splits into two corresponding to the order of the forms:
\beq
\begin{cases}
\mathcal{L}_{\xi^{k}_{Y}}\Theta^{n}_{\tilde C}= (i_{\xi^k}d+di_{\xi^k})\Theta^{n}_{\tilde C}=0,\\
\mathcal{L}_{\xi^{k}_{Y}}\Theta^{n+1}_{\tilde C}= (i_{\xi^k}d+di_{\xi^k})\Theta^{n+1}_{\tilde C}=0.
\end{cases}
\eeq

For the first condition we have, using formula (5.7) of Theorem 2;
\begin{multline}
di_{\xi^k}\Theta^{n}_{\tilde C}= -i_{\xi^k}d\Theta^{n}_{\tilde C}=-i_{\xi^k}[\tilde{d}\Theta_{\tilde C_{-}}-\Theta^{n+1}_{C}]=\\ = \omega^{1}_{C}(\xi)+\sum_{(\nu,j)\in P}F^{j}_{\nu}\omega^{\nu}_{j}(\xi^{k+1} )\eta - i_{\xi^{k}}(\Pi_{\mu}\omega^{\mu}\wedge \eta)+ Con =\\=
\omega^{\mu}(\xi)[d_{i}F^{i}_{\mu}+\lambda_{G,i}F^{i}_{\mu}-\Pi_{\mu}]+\sum_{(\nu,j)\in P}F^{j}_{\nu}\omega^{\nu}_{j}(\xi^{k+1})\eta - \Pi_{\mu}\xi^{\mu}\eta+ Con .
\end{multline}
Let now $s:X\rightarrow Y$ be a solution of the balance system $\mathcal{B_{C}}$.  Taking pullback of the last equality with respect to the $j^{k+1}_{p}s$ we see that the first and last terms in the right side vanishes and we get
\beq
j^{k}(s)^{*}di_{\xi^k}\Theta^{n}_{\tilde C}=dj^{k}(s)^{*}i_{\xi^k}\Theta^{n}_{\tilde C}=-(\Pi_{\mu}\xi^{\mu})\circ j^{k}(s) \eta+\left[\sum_{(\nu,j)\in P}F^{j}_{\mu}\omega^{\mu}_{j}(\xi^{k+1} )\right]\circ j^{k+1}(s)\eta.
\eeq
Now we recall definition of the \textbf{multimomentum mapping}, see  \cite{GIMMSY I,LM}
\begin{definition} A multimomentum mapping $J^{\hat C}:J^{k}_{p}(\pi)\rightarrow \Lambda^{n}(X)\otimes \mathfrak{g}^{*}$ of a CCR ${\hat C}$ is defined as
\beq
J^{\hat C}(z)(\xi^{k}_{Y})=-i_{\xi^{k}_{Y} }\Theta_{\hat C}^{n}(z).
\eeq
\end{definition}
\begin{lemma}
\beq
J^{\hat C}(z)(\xi^{k}_{Y})=-[(p+z^{\mu}_{j}F^{j}_{\mu})\xi^{i}+F^{i}_{\mu}\omega^{\mu}(\xi)]\eta_{i}+DCon
\eeq
where $DCon$ is the sheaf of contact forms whose differential is also contact.
\end{lemma}
\begin{proof} We have
\begin{multline*}
i_{\xi^{k}_{Y} }\Theta_{\hat C}^{n}(z)=i_{\xi}\Theta_{\hat C}^{n}(z)=p\xi^{i}\eta_{i}+F^{i}_{\mu}\xi^\mu \eta_{i}-F^{i}_{\mu}\xi^{j}dy^\mu \wedge \eta_{ij}=\\
(p\xi^{i}+\xi^\mu F^{i}_{\mu} ) \eta_{i}-F^{i}_{\mu}\xi^{j}(\omega^\mu+z^{\mu}_{k}dx^k ) \wedge \eta_{ij}=(p\xi^{i}+\xi^\mu F^{i}_{\mu} ) \eta_{i}-F^{i}_{\mu}\xi^{j}\omega^{\mu}\wedge \eta_{ij} -F^{i}_{\mu}\xi^{j}z^{\mu}_{k}(\delta^{k}_{j}\eta_{i}-\delta^{k}_{i}\eta_{j})=\\
=(p\xi^{i}+\xi^\mu F^{i}_{\mu} ) \eta_{i}-F^{i}_{\mu}\xi^{j}\omega^{\mu}\wedge \eta_{ij} -(F^{i}_{\mu}\xi^{j}z^{\mu}_{j}\eta_{i}-F^{i}_{\mu}\xi^{j}z^{\mu}_{i}\eta_{j})=\\
=[p\xi^{i}+\xi^\mu F^{i}_{\mu}-F^{i}_{\mu}\xi^{j}z^{\mu}_{j}+F^{j}_{\mu}\xi^{i}z^{\mu}_{j}]\eta_{i}-
F^{i}_{\mu}\xi^{j}\omega^{\mu}\wedge \eta_{ij}=\\=
[(p+z^{\mu}_{j}F^{j}_{\mu})\xi^{i}+F^{i}_{\mu}\omega^{\mu}(\xi)]\eta_{i}-F^{i}_{\mu}\xi^{j}\omega^{\mu}\wedge \eta_{ij}.
\end{multline*}
Now we notice that not just the last term in this equality is contact but its differential is contact as well, namely
\begin{multline*}
d[F^{i}_{\mu}\xi^{j}\omega^{\mu}\wedge \eta_{ij}]=Con -F^{i}_{\mu}\xi^{j}dz^{\mu}_{k}\wedge dx^k \wedge \eta_{ij}=Con - F^{i}_{\mu}\xi^{j}dz^{\mu}_{k}\wedge (\delta^{k}_{j}\eta_{i}-\delta^{k}_{i}\eta_{j})=\\
=Con - F^{i}_{\mu}\xi^{j}(\omega^{\mu}_{k }+z^{\mu}_{kl}dx^l )\wedge (\delta^{k}_{j}\eta_{i}-\delta^{k}_{i}\eta_{j}) =Con - F^{i}_{\mu}\xi^{j}z^{\mu}_{k l}(\delta^{k}_{j}\delta^{l}_{i}-\delta^{k}_{i}\delta^{l}_{j})\eta=\\
=Con -F^{i}_{\mu}\xi^{j}(z^{\mu}_{ij}-z^{\mu}_{ij} )=Con.
\end{multline*}
\end{proof}

Using the formula (7.4) we obtain the proof of the first statement of the next Theorem. Proof of other statements of Theorem is straightforward.
\begin{theorem}
\textbf{Noether Theorem (general).}  Let $\mathcal{C}$ be a constitutive relation defined on a partial k-jet bundle
$J^{k}_{p}(\pi)$, let $\hat C$ be an arbitrary covering CCR (with an arbitrary $p$) of $\mathcal{C}$ and $\tilde C$ - \emph{lifted} CCR of $\mathcal{C}$. Let a Lie group $G\subset
Sym(\tilde{C})\subset Aut(\pi)$ be a \emph{geometrical symmetry group of the flux part} $\Theta^{n}_{C}$ of CR $\mathcal{C}$.  Then,
\begin{enumerate}
\item
 For all $\xi  \in \mathfrak{g}$ and
for all solutions $s\in Sol(X,\mathcal{B_{C}}) $ of the balance system $\bigstar$,
\begin{multline}
d[J^{\hat C}(j^{k}(s)(x))(\xi^{k}_{Y})]=-d j^{k\ *}(s)(x)
[ ((p+z^{\mu}_{j}F^{j}_{\mu})\xi^{i}_{Y}+F^{i}_{\mu}\omega^{\mu}_{Y}(\xi ))\eta_{i}]=\\
=(\Pi_{\mu}\xi^{\mu}_{Y})\circ j^{k}(s)-
\left[\sum_{(\nu,j)\in P}F^{\nu}_{\nu}\omega^{\nu}_{j}(\xi^{k+1}_{Y} )\right]\circ j^{k+1}(s).
\end{multline}
\item If ${\hat C}={\tilde C}$, then the term $(p+z^{\mu}_{j}F^{j}_{\mu})\xi^{i}_{Y}$ in the left side vanishes.
\item
If $\mathfrak{g}_{Y} \subset \mathcal{X}_{C}(Y)$ on $Y$, last term in the right side drops out.
\item The same formula holds for a Lie group of projectable Cartan symmetries $G\subset Aut_{p}(\pi_{k0})$ of the space $J^{k}_{p}(\pi)$ and its Lie algebra $\mathfrak{g}_{Y} \subset \mathcal{X}(J^{k}_{p}(\pi)).$
\item  If $\mathfrak{g}_{Y} \subset \mathcal{X}_{C}(Y)$ and if the system $B_{C}$ is the
\emph{conservative laws system} (i.e. if $\Pi_{\mu}=0,\ \mu=1,\ldots ,m$), then $\forall \xi \in \mathfrak{g}$ and for all $s\in Sol(C)$ the \bf{Noether conservation law} holds:
 \[ d[ j^{k\ *}(s)(x)F^{i}_{\mu}\omega^{\mu}_{Y}(\xi )\eta_{i}]= ((F^{i}_{\mu}\omega^{\mu}_{Y}(\xi ))\circ j^{k\ *}(s)(x))_{;i}=0 .\]
 \end{enumerate}
 \end{theorem}
\subsection{Semi-Lagrangian and RET cases.}
In the special cases of semi-P-Lagrangian and RET constitutive relations the formulation of Noether Theorem is conveniently simplifies.
\begin{theorem}\textbf{Noether Theorem for} $C_{L,\Pi}, k=1$.  Let ${\hat C}_{L,\Pi}$ be a semi-P-Lagrangian CCR with \[
\Theta_{\hat C}=(L-z^{\mu}_{i}L_{z^{\mu}_{i}})\eta+\sum_{(\mu,i)\in P}L_{z^{\mu}_{i}}dy^\mu \wedge
\eta_{i}+\sum_{(\mu,i)\notin P}F^{i}_{\mu}dy^\mu\wedge \eta_{i}+\Pi_{\mu}dy^\mu \wedge \eta.
\]
Let $\xi \in \mathcal{X}_{p}(Y)$ be an infinitesimal variational symmetry of the CR ${\hat C}_{L,\Pi}$. Then, for all $s\in
Sol(\mathcal{C})$ of the system $\bigstar$
\begin{enumerate}
\item
\begin{multline} d [(j^{1}(s))^{*}J^{{{\hat C}_{L,\Pi}}}(z)(\xi^1)]=-[(L(j^{1}s(x))\xi^{i}+\sum_{(\mu,i)\in P}(L_{z^{\mu}_{i}}\omega^{\mu}(\xi^1))\circ j^{1}s(x) + \\ + \sum_{(\mu,i)\notin P} (F^{i}_{\mu}\omega^{\mu}(\xi^1))\circ j^{1}s(x)]_{;i} \eta  =(\omega^{\mu}(\xi^1)\Pi_{\mu})\circ j^{1}(s))^{*}\eta.
 \end{multline}
\item If $J^{1}_{p}(\pi)=J^{1}(\pi)$ is the full 1-jet bundle, last equality takes the form
\[
-[(L(j^{1}s(x))\xi^{i}+(L_{z^{\mu}_{i}}\omega^{\mu}(\xi^1))\circ j^{1}s(x)]_{;i}   =(\omega^{\mu}(\xi^1)\Pi_{\mu})\circ j^{1}(s))^{*}.
\]
 \item If, in addition,  $\Pi_{\mu}=0,\ \mu=1,\ldots ,m$, then $\forall \xi \in \mathfrak{g}$ and for all
solutions $s\in \Gamma(\pi)$ the \emph{Noether conservation law}
holds:
 \[ d [(j^{1}(s))^{*}J^{{{\hat C}_{L,\Pi}}}(z)(\xi^1)]=[(L(j^{1}s(x))\xi^{\mu}+(L_{z^{\mu}_{i}}\omega^{\mu}(\xi^1))\circ j^{1}s(x)]_{;i}\eta =0 .\]
 \end{enumerate}
\end{theorem}
Last statement of this theorem is the standard form of Noether Theorem in Lagrangian Field Theory.
\begin{theorem}
\textbf{Noether Theorem, RET,\ $k=0$}.  Let $\tilde C$ be a lifted covering constitutive relation of the RET type
with $ \Theta_{\tilde C}=F^{i}_{\mu}dy^\mu \wedge \eta_{i}+\Pi_{\mu}dy^\mu \wedge
\eta,$ and let $\xi \in \mathcal{X}(Y)$ be a variational symmetry of $\tilde C$. Then for all solutions $s\in \Gamma(\pi)$ of the balance system $\bigstar$,
\[ d [(j^{1}(s))^{*}F^{i}_{\mu}\omega^{\mu}(\xi^1 )\eta_{i})]=(\omega^{\mu}(\xi^1 )\Pi_{\mu})\circ j^{1}(s))^{*}\eta.
 \]
 \end{theorem}
\subsection{Full invariance and the source charge.}
Let now condition (7.1) is fulfilled i.e. $G$ is symmetry of \emph{both flux \textbf{and} modified source term} $\Theta^{n+1}_{C}=\Pi_{\mu}\omega^\mu \wedge \eta +F^{i}_{\mu}\omega^{\mu}_{i}\wedge \eta$ of a covering constitutive relation $\hat{C}$.
 Notice that this term depends only on the $\mathcal{C}$ but not on the lifting of the CR $\mathcal{C}$ to the covering constitutive relation.\par
  Then, for any $\xi \in \mathfrak{g}\subset \mathcal{X}(J^{k}_{p}(\pi))$, lifting $\xi$ and the Poincare-Cartan form to $J^{k+1}_{p}(\pi)$ and using the fact that any (n+1)-form on $J^{k+1}_{p}(\pi)$ is contact we have

\begin{multline} \pi^{(k+1)*}_{k}di_{\xi^k}\Theta_{\hat{C}}^{n+1}=-i_{\xi^{k+1}}d\Theta_{\hat{C}}^{n+1}= -i_{\xi^{k+1}}(d\Pi_{\mu}\wedge \omega^\mu \wedge \eta +\Pi_{\mu}d\omega^{\mu}\wedge \eta  +d(F^{i}_{\mu})\wedge \omega^{\mu}_{i}\wedge \eta +F^{i}_{\mu}d\omega^{\mu}_{i}\wedge \eta))=\\=
-i_{\xi^k}[(d_{h}\Pi_{\mu} +d_{v}\Pi_{\mu})\wedge \omega^{\mu}\wedge \eta +d_h F^{i}_{\mu}\wedge \omega^{\mu}_{i}\wedge \eta+d_v F^{i}_{\mu}\wedge \omega^{\mu}_{i}\wedge \eta]=\\=
-i_{\xi^k}[(d_{i}\Pi_{\mu}dx^i +d_{v}\Pi_{\mu})\wedge \omega^{\mu}\wedge \eta +d_i F^{i}_{\mu}dx^i \wedge \omega^{\mu}_{i}\wedge \eta+d_v F^{i}_{\mu}\wedge \omega^{\mu}_{i}\wedge \eta]=\\ =-i_{\xi^k}[d_{v}\Pi_{\mu})\wedge \omega^{\mu}\wedge \eta+d_v F^{i}_{\mu}\wedge \omega^{\mu}_{i}\wedge \eta]=-i_{\xi^k}2Con =  Con.
\end{multline}

Here, as before $Con$ is a contact form while 2Con - two-contact form (see Sec.2).  We have used here the equalities $d\omega^\mu \wedge \eta =-dz^{\mu}_{ij}\wedge dx^j \wedge \eta =0$ and the same for $d\omega^{\mu}_{i}\wedge \eta$. During this calculation we repeatedly used the equality $dx^j \wedge
\eta =0$ \emph{but we have not used the balance system}.

Reminding the decomposition $d=d_{h}+d_{v}$ as the sum of vertical and horizontal differentials we see that the (n+1)-form
\beq Q(\xi)= i_{\xi^k}\Theta_{\hat{C}}^{n+1}=i_{\xi^k}[\Pi_{\mu}\omega^\mu \wedge \eta +F^{i}_{\mu}\omega^{\mu}_{i}\wedge \eta]= i_{\xi^k}K_{\mathcal{C}}\eeq
has contact differential and therefore, \textbf{zero horizontal differential}: $d_{h}Q(\xi)=0.$  Thus, it defines the \textbf{class of cohomology } $[Q(\xi)]$ of the \emph{horizontal complex} $(\bigwedge^{*}(J^{\infty }_{p}(\pi)),d_{h})$:
\[
[Q(\xi)=i_{\xi^k}K_{\mathcal{C}}]\in H^{n+1}_{hor}(J^{\infty }_{p}(\pi)))
\]
If the class $[Q]$ is zero, then $Q(\xi)=d_{h}\Phi(\xi)+Con$ is the sum of horizontal differential of a form $\Phi (\xi)$ and a contact form and, therefore is also the sum of a differential $d\Phi (\xi)$ and the contact form \[Q(\xi)=d\Phi (\xi) +Con.\]\par

  Applying now the pullback by $j^{k}(s)$ (and $j^{k+1}(s)$ where appropriate) and recalling that we have not used the balance equation in the calculation above, we prove the following

\begin{theorem} Let, in addition to the conditions of Theorem 8,  $\mathfrak{g}$ is the Lie algebra of the variational (infinitesimal) symmetries of the source part
of the constitutive relation, i.e. (7.2) is true. Then
\begin{enumerate}
\item
\[
di_{\hat{\xi}}\Theta^{n+1}_{\hat{C}}=\text{Cont}\Rightarrow
dj^{k}(s)^{*}i_{\hat{\xi}}\Theta^{n+1}_{\hat{C}}=0
\]
 for all sections $s$ of the configurational bundle $\pi$.
 \item Form $Q(\xi)=\pi^{(k+1)*}_{k} i_{\xi^k}\Theta_{\hat{C}}^{n+1}=i_{\xi^k}K_{\mathcal{C}}$ defines the \textbf{class of cohomology} $[Q(\xi)]$ of the horizontal complex $(\bigwedge^{*}(J^{\infty}_{p}(\pi)),d_{h}),$ \cite{KV} - $\mathfrak{g}$-\textbf{charge of the source} $\tilde{\Pi}=(\Pi_{\mu}\omega^\mu +F^{i}_{\mu}\omega^{\mu}_{i})\wedge \eta $.

\item If the class $[Q(\xi)]$ is zero for all $\hat{\xi} \in \mathfrak{g}$, there exists a form $\Phi_{\mathcal{C}}(\hat{\xi})$ \emph{linearly depending on} $\hat{\xi}$ (and its derivatives) such that $Q(\hat{\xi})=d\Phi_{C}(\hat{\xi}) +Con$ is the sum of a differential $d(\Phi_{\mathcal{C}}(\hat{\xi}))$ and the contact form.  In this case locally (and in a topologically trivial domain, globally)
\[
j^{k}(s)^{*}i_{\hat{\xi}}\Theta^{n+1}_{\hat{C}}=dj^{k+1}(s)^{*}\Phi_{\mathcal{C}}(\hat{\xi} )
\]
for this  $\mathfrak{g}^*$-valued n-form $\Phi_{\mathcal{C}}$ (that is natural to call the $\mathfrak{g}$-\textbf{potential of the source}  $\Theta^{n+1}_{C}$).
\end{enumerate}
\end{theorem}

\begin{corollary} If $G$ is the Lie group of symmetries of a constitutive relation $C$ and if the class $[Q]$ in the last Proposition is trivial, then there exists the (locally defined) $\mathfrak{g}^*$-valued n-form  $\Phi_{\mathcal{C}}:\xi\rightarrow \Lambda^{n}(J^{k}_{p}(\pi))$ such that in the conditions of Theorem 11 the following \textbf{conservation law} will holds
\[
d[J^{\widehat{\mathcal{C}}}(j^{k}(s)(x)(\xi)-j^{k+1}(s)^{*}\Phi_{\mathcal{C}}(\hat{\xi}) ]=0
\]
for all solutions $s\in \Gamma(\pi)$ of the balance system $\mathcal{B_{C}}$ and all $\hat{\xi} \in \mathfrak{g}$.
\end{corollary}

\subsection{Examples: Energy-Momentum Balance law, case of gauge symmetries.}

\begin{example}\textbf{ Energy-Momentum Balance Law, lifted CCR.}
Let $\nu$ be a connection in the bundle $\pi:Y\rightarrow X$ with the connection form
$K_{\nu}=\partial_{y^\mu}\otimes (dy^{\mu}-\Gamma^{\mu}_{i}dx^i ).$\par
   Then, the horizontal lift of the basic vector field  $\partial_{x^i}$ in $Y$  is $\xi_{i}=\partial_{x^i }+\Gamma^{\mu}_{i}\partial_{y^\mu}$.
   Its lift to the Lie vector field in $J^{k}_{p}(\pi)$ has the form
   \[\xi^{k}_{i}=\partial_{x^i}+ \Gamma^{\mu}_{i}\partial_{y^\mu}+\sum_{(\mu,j)\in P}d_{j}\Gamma^{\mu}_{i}\partial_{z^{\mu}_{j}}+\ldots .\] \par
\begin{remark} Connection $\nu$ is assumed to be compatible with the partial structure of $J^{1}_{p}(\pi)$. In a case of $K\oplus K'$-structure this means that the $\nu$-horizontal lift $\xi $ to $Y$ of any vector field $\bar{\xi} \in \mathcal{X}(X)$ preserving the $K\oplus K'$-structure should be such that its flow prolongation to $J^{1}_{p}(\pi)$ is possible and preserves the (partial ) Cartan structure. In the case of an integrable AP-structure this requires that the components $\Gamma^{\mu}_{i}$ of the connection $\nu$ depends on the variables $x^j, \partial_j \in K$ and on $y^\mu$ (see \cite{Pr1}, Sec.8.2) but not on the complemental variables in $X$.\end{remark}
Consider a $\nu$-homogeneous constitutive law $\mathcal{C}$ and the corresponding balance system $\mathcal{B_{C}}$. Let $\tilde C$ be the lifted CCR corresponding to $\mathcal{C}$ and $\Theta_{\tilde C_{-}}=F^{i}_{\mu}\omega^{\mu}\wedge \eta_{i}-\Pi_{\mu}dy^\mu \wedge \eta- F^{i}_{\mu}\omega^{\mu}_{i}\wedge \eta$ - corresponding (corrected) Poincare-Cartan form.
 Calculate now
\[
i_{\hat{\partial}^{1}_{i}}\Theta_{\tilde C }^{n}=i_{\hat{\partial}^{1}_{i}}F^{k}_{\mu} \omega^\mu \wedge \eta_{k}= F^{k}_{\mu}(\Gamma^{\mu}_{i }-z^{\mu}_{i})\eta_{\sigma} -F^{\sigma}_{i}\omega^\mu \wedge \eta_{ki}.
\]
Next one has,
\[
di_{\hat{\partial}^{1}_{i}}\Theta_{\tilde C }^{n}=d_{k}(F^{k}_{\mu}(\Gamma^{\mu}_{i }-z^{\mu}_{i}))\eta +F^{k}_{\mu}(\Gamma^{\mu}_{i }-z^{\mu}_{i})\lambda_{G,k}\eta -F^{k}_{\mu}d\omega^{\mu}\wedge \eta_{ki}+Con=(d_{k}+\lambda_{G,k})[F^{k}_{\mu}(\Gamma^{\mu}_{k }-z^{\mu}_{k})]\eta +con,
\]

since $j(s)^{*} d\omega^{\mu}=dj^{*}s\omega^\mu =0$.\par

For the right side of balance law (7.7) we have

\[
\Pi_{\mu}\omega^\mu (\hat{\partial}^{1}_{i}) \eta +F^{k}_{\mu}\omega^{\mu}_{i}(\xi^2 )\eta=[\Pi_{\mu}(\Gamma^{\mu}_{i}-z^{\mu}_{i})+F^{k}_{\mu}(d_{k}\Gamma^{\mu}_{i}-z^{\mu}_{ki})]\eta ,
\]
since $\omega^{\mu}({\hat \xi}_{i})=\Gamma^{\mu}_{i}-z^{\mu}_{i}$. \par
Applying the pullback by the $j^{2}s$ we get for the right side expression
\[
[\Pi_{\mu}(j^{1}s) (\Gamma^{\mu}_{i}-z^{\mu}_{i})\circ j^{1}(s) +F^{k}_{\mu}\circ j^{1}s (d_{k}\Gamma^{\mu}_{i} \circ j^{2}s-s^{\mu}_{,ki })] \eta .
\]

  As a result, the balance law (7.7) takes, for the vector field ${\hat \xi}_{i}$, the form

\beq (d_{k}+\lambda_{G,k})[F^{k}_{\mu}(\Gamma^{\mu}_{k }-z^{\mu}_{k})]\circ j^{2}s=
\Pi_{\mu}(j^{1}s) (\Gamma^{\mu}_{k}-z^{\mu}_{k})\circ j^{1}(s)  +F^{k}_{\mu}\circ j^{1}s (d_{k}\Gamma^{\mu}_{i} \circ j^{2}s-s^{\mu}_{,ki }).\eeq

Thus, the Energy-Momentum Tensor for the constitutive relation $\mathcal{C}$ has, thus, the form
\beq
T^{j}_{i}=F^{j}_{\mu}(\Gamma^{\mu}_{i }-z^{\mu}_{i}).
\eeq
In the left side of (7.11) stays the covariant divergence of the energy-momentum tensor.
Right side of (7.11) represents the \textbf{source/dissipation term}.
\par
Calculating derivative in the left side of (31.17) and canceling similar terms we reduce it to the equality
\begin{multline}
[(d_{k}+\lambda_{G,k})F^{k}_{\mu}](\Gamma^{\mu}_{i }-z^{\mu}_{i})]\circ j^{2}s=
\Pi_{\mu}(j^{1}s) (\Gamma^{\mu}_{i}-z^{\mu}_{i})\circ j^{1}(s)  \Leftrightarrow \\ \Leftrightarrow [(d_{k}+\lambda_{G,k})F^{k}_{\mu}-\Pi_{\mu}](\Gamma^{\mu}_{i }-z^{\mu}_{i})]\circ j^{2}s=0.
\end{multline}
This gives the explicit presentation of the components of the energy-momentum balance law as the linear combination of the original balance laws with variable coefficients $(\Gamma^{\mu}_{i }-z^{\mu}_{i}).$
\end{example}

\begin{example}\textbf{Energy-Momentum balance law, semi-P-Lagrangian case.}
Let $C_{L,\Pi}$ be a semi-P-Lagrangian covering constitutive relation (see Theorem 3).  Let us specialize the Noether balance law given in Theorem 8 above for the case of the vector fields $\xi_{i}=\partial_{i}+\Gamma^{\mu}_{i}\partial^\mu$ introduced in the previous Example.  This balance law takes, for a solution $s:X\rightarrow Y$ of the balance system $\bigstar$, the form
\beq
[(L\delta^{j}_{ii}+\sum_{(\mu,j)\in P}L_{,z^{\mu}_{j}}\cdot (\Gamma^{\mu}_{i}-z^{\mu}_{i})+
\sum_{(\mu,j)\notin P}F^{i}_{\mu}\cdot (\Gamma^{\mu}_{i}-z^{\mu}_{i}))\circ j^{1}(s)(x)]_{;i}=-(\Pi_{\mu}\cdot (\Gamma^{\mu}_{i}-z^{\mu}_{i}))\circ j^{1}(s)(x).
\eeq
In particular, for the case of the full 1-jet bundle $J^{1}(\pi)$ and of zero connection $\Gamma^{\mu}_{i}\equiv 0$ this balance law takes the conventional  form
\beq
[(L(j^{1}(s)(x))\delta^{j}_{i}-L_{,z^{\mu}_{j}}(j^{1}(s)(x)) s^{\mu}_{,i})]_{;i}=(\Pi_{i}z^{\mu}_{i})(j^{1}(s)(x)).
\eeq
\end{example}
\begin{example} \textbf{Pure gauge symmetry transformation.}
Let $\xi=\xi^{\mu}\partial_{y^\mu}$ be a vertical (pure gauge) symmetry transformation of a constitutive
relation $\mathcal{C}$. Then,  the Noether balance law corresponding to the vector field  $\xi$ has the form
\beq [(\xi^{\mu}F^{i}_{\mu})( j^{1}_{p}(s))]_{;x^i}= (\xi^{\mu}\Pi_{\mu})( j^{1}_{p}(s))  \eeq
- \emph{the secondary balance law} defined by the vector field $\xi \in g(C)$, see Lemma 9 below.
\end{example}

\section{ Secondary balance laws and the entropy principle.}
 Let the system of PDE ($\bigstar$) be a balance system corresponding to a constitutive relation $\mathcal{C}$. A natural question that generalizing the "entropy principle" of Continuum Thermodynamics (\cite{MR}) is - are there, except of the linear combinations of balance equations of the system  $\star$, \emph{nontrivial} (see below) balance laws defined on the same bundle $J^{1}_{p}(\pi)$ where the CR $\mathcal{C}$ is defined \emph{that are satisfied by all the solutions of  the balance system} $\star$:

\begin{definition} Let $\mathcal{C}$ be a constitutive relation defined on a partial k-jet bundle $J^{k}_{p}(\pi)$ and let $\mathcal{B_{C}}$ be corresponding balance system ($\bigstar$). We call a balance law
  \beq (K^{i}\circ j^{q}_{p}(s))_{,i}=Q\circ j^{q+1}_{p}(s). \eeq \textbf{of the order q} i.e. given by a (n+1)+(n+2)-form $\sigma = K^{i}\eta_{i}+Q\eta$ on the space $J^{q}_{p}(\pi)+J^{q+1}_{p}(\pi)$ - the \textbf{secondary balance law} for the system $\bigstar$ if \textbf{any} solution $s:X\rightarrow Y$ of the balance system ($\bigstar$) is at the same time solution of the balance law (8.1).
\end{definition}
Following is the list of four classes of secondary balance laws of the balance systems.\par
\begin{enumerate}
\item An interesting class of the secondary balance laws, including the linear combinations of the balance laws of the system ($\bigstar$), is determined by the following

\begin{lemma} Let a vertical vector field $\xi =\xi^\mu \partial_{y^\mu}\in V(\pi)$ is such that
  the condition $FDiv (\xi)=F^{\mu}_{i}d_i \xi^{\mu}=0$ is fulfilled.
  Then the balance law
\[
j^{1\ *}_{p}(s)d(\xi^{\mu}F^{\mu}_{i}\eta_{i})=\xi^{\mu}\Pi_{\mu}\eta \Leftrightarrow
((\xi^{\mu}F^{\mu}_{i})\circ j^{1}(s))_{,x^i}=(\xi^{\mu}\Pi_{\mu})\circ j^{1}(s)
\]
belongs to the space $\mathcal{BL}_{\mathcal{C}}.$
\end{lemma}
\begin{proof} Follows from $d(j^{1\ *}_{p}(s)\xi^{\mu}F^{\mu}_{i}\eta_{i})=
\xi^{\mu}d(j^{1\ *}_{p}(s)F^{\mu}_{i}\eta_{i})+j^{1\ *}_{p}(s)FDiv(\xi)\eta .$
\end{proof}

\item If a Lie group $G$ is the (geometrical) symmetry group of the balance system $\bigstar$ , it determines the family of the balance laws
corresponding to the elements of Lie algebra $\mathfrak{g}$, see previous section. If the second order source part of these balance laws vanishes, we get the subspace $\mathcal{BL}_\mathcal{C,\mathfrak{g}}$ of the space $ \mathcal{BL}_\mathcal{C}.$
\item
The \textbf{entropy principle} of Thermodynamics (see (\cite{ME,MR,Mu}) requires that the entropy balance
\beq h^{i}_{,i}=\Sigma, \eeq
 with the entropy density $h^0$, entropy flux $h^A,A=1,2,3$, and entropy
production plus source $\Sigma=\Sigma_{s}+\Sigma_{p} $ belongs to the space $\mathcal{BL}_{\mathcal{C}}$ of the balance system of a given
theory. This requirement place a serious restrictions on the form of constitutive relation $\mathcal{C}$
and leads to the construction of a dual system in terms of Lagrange-Liu fields $\lambda^\mu $ considered below
(see \cite{MR}).  Even more serious constitutional restriction is the proper II law of thermodynamics requiring that the entropy production $\Sigma_p$ is nonnegative, \cite{MR,Mu}.\par

\item If an integrable dynamical system can be formulated as a system of balance equations (KdV equation is an example) then all the higher order conservation laws are the secondary balance laws of this balance system.
\end{enumerate}

\subsection{Secondary balance laws for the RET balance systems}
As an illustration of the geometrical approach to the balance systems and the example of application we present here some results from the paper \cite{Pr2} (in preparation).
In the Rational Extended Thermodynamics developed by I.Muller, T.Ruggeri, and I. Shish-Liu, \cite{MR} the space of the dynamical fields $y^\mu$ is chosen to be large enough so that the constitutive relation $C$ is defined on the bundle $Y$ and, in our terms, is represented by the section of the bundle $\Lambda^{n+(n+1)}_{2/1}(Y)\rightarrow Y$. \par A constitutive relation $\mathcal{C}$ is called regular if the matrix $\frac{\partial F^{0}_{\mu}}{\partial y^\mu}$ is \emph{nondegenerate}. For a regular CR $\mathcal{C}$ one can introduce new field variables $w^\mu=F^{0}_{\mu}$ and to write the balance system in the form $\partial_{t}w^\mu +\partial_{x^i}F^{\mu}_{i}=\Pi_\mu,\ \mu=1,\ldots, m$.

\begin{theorem}
\item Let $\mathcal{C}$ be a a constitutive relation of the RET type with $F^{0}_{\mu}=y^\mu$.
\begin{enumerate}
\item There is a bijection between
\begin{enumerate}
\item Functions $h^{0}(y^\mu)$ satisfying to the (overdetermined) system of PDE
\beq
(\eta^{A}_{\sigma}\partial_{y^\nu}-\eta^{A}_{\nu}\partial_{y^\sigma})h^0 =0,\ \sigma \ne \nu,A=1,2,\ldots ,n,
\eeq
where $\eta^{A}_{\sigma }=F^{A}_{\mu,y^\sigma }(x,y)\partial_{y^\mu}$ and such that the (\emph{vertical}) Hessian
\[ Hess(h^{0})(x_{0},y_{0})=\left( \frac{\partial^2 h^0}{\partial y^{\mu}\partial
y^{\nu}}\right)\vert_{x=x_{0},y=y_{0}}\] is nondegenerate in a neighborhood of a point $(x_{0},y_{0})$, and
\item Secondary balance laws () with the functionally independent Lagrange-Liu multipliers $\lambda^\mu =\frac{\partial h^0}{\partial y^\mu}$
in a neighborhood of a point $(x_{0},y_{0})$.
\end{enumerate}
\item In terms of local vertical variables $\lambda^\mu$ the density, flux and the source of the corresponding secondary balance law are
\beq
\begin{cases}
K^0 (x,y)=h^{0}(\lambda(y)),\\
K^\nu (x,y)=\frac{h^{0}(\lambda)}{\partial \lambda^\nu} (\lambda(y)),\ \nu =1,2,3,\\
Q(x,y)= \lambda^\mu (y)\Pi_\mu (x,y).
\end{cases}
\eeq
\end{enumerate}
\end{theorem}

\subsection{Example: Cattaneo heat propagation law.}

Balance equations of this model have the form
\beq
\begin{cases}
\partial_{t}(\rho \epsilon)+div(q)=0,\\
\partial_{t}(\tau q)+ \nabla \Lambda(\vartheta) =-q.
\end{cases}
\eeq
Since $\rho$ is not considered here as a dynamical variable, we merge it with the field $\epsilon$ and from now on and till the end it will be omitted. On the other hand, in this model the the energy $\epsilon$ depends on temperature $\vartheta$ \emph{and on the heat flux} $q$ (see \cite{JCL}, Sec.2.1.2 or, by change of variables, temperature $\vartheta=\vartheta(\epsilon, q)$ will be considered as the function of dynamical variables.\par

The secondary balance laws for Cattaneo model (including the original balance laws) have the form
\begin{multline}
\begin{pmatrix}K^0\\K^1\\ K^2\\ K^3 \\ Q\end{pmatrix} = a^0 \begin{pmatrix} \e \\ q^1\\q^2\\q^3 \\0
\end{pmatrix}+ \sum_{A}k^A \begin{pmatrix} \tau(\va )q^A \\ \delta^{1}_{A} \Lambda( \va) \\ \delta^{2}_{A} \Lambda(\va) \\ \delta^{3}_{A} \Lambda(\va) \\ - q^A\end{pmatrix}+\begin{pmatrix} f_0\\ m^1\\m^2\\m^3 \\0
\end{pmatrix}+ \\ + \alpha \begin{pmatrix}\hat{\lambda}^0 \e -\int^{\va}\lambda^{0}_{,\va} \e^{eq} ds +\tau(\va )\Lambda^{-1}_{\va}[\frac{1}{2}\lambda^{0}_{\va}\Vert q\Vert^2+\hat{K}^{A}_{,\va}(\va)q^A]\\ \hat{\lambda}^{0}(\va)q^1 +\hat{K}^{1}(\va)\\ \hat{\lambda}^{0}(\va)q^2 +\hat{K}^{2}(\va)\\ \hat{\lambda}^{0}(\va)q^3 +\hat{K}^{3}(\va)\\ -\Lambda^{-1}_{,\va }(\hat{\lambda}^{0}_{\va}\Vert q\Vert^2 +\hat{K}^{A}_{,\va}(\va)q^A)\end{pmatrix}.
\end{multline}
First and second balance laws in the system (8.6) are the balance laws of the original Cattaneo system. Third one one is the trivial balance law (see Sec.2).\par

Restrictions placed by the system (8.3) leads to the expression of internal energy
\begin{multline}
\e =\e^{eq}(\va)+\frac{\tau_{,\va}}{2\Lambda_{,\va}}\Vert q\Vert^2 -\frac{\tau(\va)}{\lambda^{0}_{,\va}}\left[ \frac{1}{2}\left(\frac{\lambda^{0}_{,\va}}{\Lambda_{,\va}} \right)_{,\va}\Vert q\Vert^2+ \left(\frac{{\tilde K}^{A}_{,\va}}{\Lambda_{,\va}}\right)_{,\va}q^A \right].
\end{multline}

  Forth column gives the balance law with the production term
\begin{multline}
 -\Lambda^{-1}_{,\va }(\hat{\lambda}^{0}_{\va}\Vert q\Vert^2 +\hat{K}^{A}_{,\va}(\va)q^A)=
 -\Lambda^{-1}_{,\va }\hat{\lambda}^{0}_{\va}\left[\sum_{A}( q^A +\frac{\hat{K}^{A}_{,\va}(\va)}{2\hat{\lambda}^{0}_{\va}})^2 -\sum_{A} \left( \frac{\hat{K}^{A}_{,\va}(\va)}{2\hat{\lambda}^{0}_{\va}}\right)^2\right]
\end{multline}
For a fixed $\va$ this expression may have constant sign as the function of $q^A$ if and only if all $\hat{K}^{A}_{,\va}(\va)=0$. Therefore this is possible only if
the internal energy (8.7) has the form
\beq
\e=\e^{eq}(\va)+\left[ \frac{\tau_{,\va}}{2\Lambda_{,\va}} -\frac{\tau(\va)}{2\hat{\lambda}^{0}_{,\va}} \left(\frac{\hat{\lambda}^{0}_{,\va}}{\Lambda_{,\va}} \right)_{,\va} \right] \Vert q\Vert^2
\eeq
with some function ${\hat \lambda}^{0}(\va)$. As a result, Cattaneo model has the secondary balance law
\begin{multline}
\partial_{t}\left[ \hat{\lambda}^0 \e -\int^{\va}\hat{\lambda}^{0}_{,\va} \e^{eq}ds +\frac{1}{2}\tau(\va )\Lambda^{-1}_{\va}\hat{\lambda}^{0}_{\va}\Vert q\Vert^2\right] +\partial_{x^A}\left[\hat{\lambda}^{0}(\va)q^A \right]= -\Lambda^{-1}_{,\va }\hat{\lambda}^{0}_{\va}\Vert q\Vert^2.
\end{multline}
 \textbf{with the production term that may have constant sign - nonnegative, provided}
 \beq
 \Lambda^{-1}_{,\va }\hat{\lambda}^{0}_{\va}\leqq 0.
 \eeq
\textbf{Last inequality is the II law of thermodynamics for Cattaneo heat propagation model}.

\section{Conclusion}
In this work we've presented a variational theory of system of balance equations, realization of the constitutive relations of such system as an abstract Legendre transformation, invariant form of the balance system and the Noether Theorem associating with the geometrical symmetries of the constitutive relation the corresponding balance law (and in the appropriate cases - the conservation law). Further development of this scheme including algebraic and geometrical structures related to a balance laws, some functorial properties of developed formalism, more detailed relations to the variational bicomplex will be presented elsewhere.  Applications to the thermodynamical systems: 5-field thermodynamical system, 13-fields systems, description of secondary balance laws using the exterior differential systems (see \cite{BGG})  will be presented in the forthcoming paper \cite{Pr2}.

\section{Appendix 1. Proof of Proposition 6.}
\begin{proof} Let $\xi =\xi^i \partial_{i}+\xi^\mu \partial_{\mu}+\xi^{i}_{\mu}\partial_{z^{\mu}_{i}}$ be any vector field in $J^{1}(\pi)$. Present any vector field in the space $\Lambda^{n+1}_{2}(J^{1}(\pi))$ in the form
\[
\xi^{*} =\xi^i \partial_i+\xi^\mu\partial_\mu+\xi^{i}_{\mu}\partial_{z^{\mu}_{i}}+\xi^{q_\mu}\partial_{q_\mu}+\xi^{q^{i}_{\mu}}\partial_{q^{i}_{\mu}}.
\]
We have:
\begin{multline}
\mathcal{L}_{\xi^*}Q^{n+1}=(di_{\xi^{*}}+i_{\xi^{*}}d )[(q_\mu dy^\mu +q^{i}_{\mu}dz^{\mu}_{i})\wedge \eta]=\\
 =d[(q_\mu \xi^\mu +q^{i}_{\mu}\xi^{\mu}_{i})\wedge \eta -(q_\mu dy^\mu +q^{i}_{\mu}dz^{\mu}_{i})\wedge \xi^{i} \eta_{i}]+i_{\xi^{*}}[(dq_\mu \wedge dy^\mu +dq^{i}_{\mu}\wedge dz^{\mu}_{i})\wedge \eta]=\\
=\xi^\mu dq_\mu\wedge \eta +q_\mu d\xi^\mu\wedge \eta +\xi^{\mu}_{i}dq^{i}_{\mu} \wedge \eta +q^{i}_{\mu}d\xi^{\mu}_{i} \wedge \eta -\\
-(dq_\mu \wedge dy^\mu +dq^{i}_{\mu}\wedge dz^{\mu}_{i})\wedge \xi^{i} \eta_{i}+(q_\mu dy^\mu +q^{i}_{\mu}dz^{\mu}_{i})\wedge (\xi^{i}_{,i}\eta+\xi^i \lambda_{G,i} \eta)]+\\
 +i_{\xi^{*}}[(dq_\mu \wedge dy^\mu +dq^{i}_{\mu}\wedge dz^{\mu}_{i})\wedge \eta  ]=\\=
 \xi^\mu dq_\mu\wedge \eta +q_\mu d\xi^\mu \wedge \eta +\xi^{\mu}_{i}dq^{i}_{\mu} \wedge \eta +q^{i}_{\mu}d\xi^{\mu}_{i} \wedge \eta -\\-(dq_\mu\wedge dy^\mu +dq^{i}_{\mu}\wedge dz^{\mu}_{i})\wedge \xi^{i} \eta_{i}+(q_i dy^\mu+q^{i}_{\mu}dz^{\mu}_{i})\wedge div_{G}(\bar \xi )\eta)]+\\+
[(\xi^{q_\mu} dy^\mu +\xi^{q^{i}_{\mu}} dz^{\mu}_{i})\wedge \eta  ]+
[(-\xi^\mu dq_\mu  -\xi^{\mu}_{i}dq^{i}_{\mu} )\wedge \eta ] +
[(dq_\mu \wedge dy^\mu +dq^{i}_{\mu}\wedge dz^{\mu}_{i})\wedge \xi^{i}\eta_{i}].
\end{multline}
\par
It is easy to see that all terms containing $dq_\mu$ and  $dq^{i}_{\mu}$ will cancel and we get
\begin{multline}
\mathcal{L}_{\xi^*}Q^{n+1}=
 q_\mu d\xi^\mu \wedge \eta  +q^{i}_{\mu}d\xi^{\mu}_{i} \wedge \eta +div_{G}(\bar \xi )(q_\mu dy^\mu +q^{i}_{\mu}dz^{\mu}_{i})\wedge \eta)]+\\+
[(\xi^{q_\mu} dy^\mu +\xi^{q^{i}_{\mu}} dz^{\mu}_{i})\wedge \eta  ].
\end{multline}

Calculating differentials, leaving only vertical differential due to the presence of $\eta$ as common factor and equating result to zero we finally get
\beq
\mathcal{L}_{\xi^*}Q^{n+1}= [q_\mu d_v\xi^\mu  +q^{i}_{\mu}d_v\xi^{\mu}_{i}
 +div_{G}(\bar \xi )(q_\mu dy^\mu  +q^{i}_{\mu}dz^{\mu}_{i})+
(\xi^{q_\mu} dy^\mu +\xi^{q^{i}_{\mu}} dz^{\mu}_{i}) ]\wedge \eta =0
\eeq
Using the fact that the $\xi$ is projectable to $X$, so that $\xi^\mu$ depend only on $X$ and using relations
\[
d_v f=f_{,y^\nu}\omega^j+f_{,z^{\nu}_{j}}\omega^{\nu}_{j}
\]
we reduce last relation to the following
\begin{multline} q_\mu (\xi^{\mu}_{y^\nu}\omega^j +\xi^{\mu}_{,z^{\nu}_{j}}\omega^{\nu}_{j} ) +q^{i}_{\mu}(\xi^{\mu}_{i,y^\nu}\omega^\nu +\xi^{\mu}_{\mu,z^{\nu}_{j}}\omega^{\nu}_{j})
 +div_{G}(\bar \xi )(q_\mu\omega^\mu  +q^{i}_{\mu}\omega^{\mu}_{i})+
(\xi^{q_\mu} \omega^\mu +\xi^{q^{i}_{\mu}} \omega^{\mu}_{i}) ]\wedge \eta =\\
=[q_\mu \xi^{\mu}_{y^\nu}+q^{i}_{\mu}\xi^{\mu}_{i,y^\nu}+div_{G}(\bar \xi )q_\mu +\xi^{q_\nu}]\omega^j \wedge \eta +[q_\mu \xi^{\mu}_{,z^{\nu}_{j}} +q^{i}_{\mu}\xi^{\mu}_{i,z^{\nu}_{j}}
 +div_{G}(\bar \xi )q^{i}_{\mu}+\xi^{q^{i}_{\mu}} ]\omega^{\nu}_{j} \wedge \eta =0
\end{multline}
From this it follows that the prolongation of $\xi$ to the $\Lambda^{n+1}_{2}(J^{1}_{p}(\pi))$ with the properties listed in the Proposition exists, is unique and given by
\beq
\begin{cases}
\xi^{q_\nu}=-q_\mu \xi^{\mu}_{y^\nu}-q^{i}_{\mu}\xi^{\mu}_{i,y^\nu}-q_\nu div_{G}(\bar \xi ) ,\\
\xi^{q^{i}_{\nu}}=-q_\mu \xi^{\mu}_{,z^{\nu}_{j}} -q^{i}_{\mu}\xi^{\mu}_{i,z^{\nu}_{j}}-q^{i}_{\nu} div_{G}(\bar \xi ).
\end{cases}
\eeq
\end{proof}

\end{document}